\documentclass{article}
\begin{document}

\title{Reduced density matrices of oscillator systems}
\author{V.V.Dodonov, O.V.Man'ko, V.I.Man'ko}
\date{}
\maketitle
\def\theequation{\thesection.\arabic{equation}}

\begin{abstract}

We study the evolution of an oscillator interacting
via the most general bilinear coupling (with time-independent coefficients)
with an ``environment'' consisting of a set of other harmonic oscillators.
We are mainly interested in a possibility of using the Fokker-Planck
equation to describe this evolution. Studying different
interaction Hamiltonians, we show that unambiguous reduction
to the Fokker-Planck equation is possible only within the framework of the
so called rotating-wave approximation. As special cases we consider
in detail the evolution of two coupled oscillators and relaxation of a
charged oscillator in a uniform magnetic field.

\end{abstract}

\setcounter{section}{0}
\setcounter{equation}{0}
\section{Statement of the problem}
\label{sec1}

This part is devoted to the multidimensional
generalizations of the harmonic oscillator model. As a rule, these are
various systems of interacting oscillators,  some of which may be placed
in external uniform electric and magnetic fields. More precisely this
class of systems can be characterized by a Hamiltonian that is assumed
to be in the form of the general inhomogeneous quadratic form
\begin{equation}\widehat {H}=\frac 12\widehat {{\bf q}}{\bf B}\widehat {{\bf q}}
+{\bf C}\widehat {{\bf q}}.\label{1.1}\end{equation}
Here, ${\bf q}$ is conceived as a $2N$-dimensional vector
($N$ being the number
of degrees of freedom) whose components are linear combinations of
Cartesian coordinates and momenta conjugated to them. The most frequent
choice is ${\bf q}=(p_1,\ldots,p_N,x_1,\ldots,x_N)$, but in the presence
of an external magnetic field it is more convenient to deal not with the
canonical momenta $p_j$ but with the kinetic momenta
$\pi_j=p_j-eA_j({\bf x})/c$, where ${\bf A}({\bf x})$ is the vector
potential. Other choices are also possible, taking into account the
concrete physical applications. For instance, the components of the vector
${\bf q}$ may be bosonic annihilation and creation operators constructed
from the coordinate and momentum operators, etc. ${\bf B}$ is a symmetric
$2N\times2N$ matrix, which may depend on time, as also the
$2N$-dimensional vector ${\bf C}$.

Multidimensional systems with Hamiltonian (\ref{1.1}) were the subject
of investigation in numerous papers: see, e.g., \cite{7,8},
\cite{13}-\cite{15} and the references therein. In the present paper we
consider the case where the multidimensional vector ${\bf q}$ may be
split in two parts: ${\bf q}=({\bf Q},\mbox{\boldmath$\xi$})$, where
the vector ${\bf Q}$ describes a small subsystem, whereas the vector
{\boldmath$\xi$} is related to a ``thermostat''. In the simplest cases
the evolution of a unidimensional harmonic oscillator coupled with a
``thermostat'' was studied, e.g., in \cite{Dekker},\cite{16}-\cite{O'Con}.
We investigate the most general quadratic ``interaction
Hamiltonians'' and compare the results of this ``microscopic'' approach
with different phenomenological models considered in reviews
\cite{Dekker,30}. Emphasis is placed on application to the problem
of harmonic oscillator relaxation, including the case where a
(charged) oscillator is placed in a uniform magnetic field.

\setcounter{equation}{0}
\section{Phenomenological Fokker-Planck equation}
\label{sec2}

The evolution of any closed system is governed by the
quantum Liouville equation for the statistical operator $\widehat{\rho}$,

\begin{equation}i\hbar\partial\widehat{\rho }/\partial t=\widehat {H}\widehat{
\rho }-\widehat{\rho}\widehat {H}.\label{2.1}\end{equation}
For any quadratic Hamiltonian (\ref{1.1})
this equation results in the {\em linear\/} equation for the average values
of the $2N$-vector $\widehat {{\bf q}}$:

\begin{equation}\langle\dot {{\bf q}}\rangle =-\Sigma {\bf B}\langle
{\bf q}\rangle -\Sigma {\bf C},\label{2.2}\end{equation}
where $\langle {\bf q}\rangle =\mbox{Tr}(\widehat{\rho}\widehat {{\bf q}}
)$, and antisymmetric nondegenerate
$2N\times 2N$ matrix $\Sigma$ with $c$-number coefficients is defined via the
commutation relations between the operators $\widehat {q}_{\alpha}$,

\begin{equation}\left[\widehat q_{\alpha},\widehat q_{\beta}\right]=-i\hbar
\Sigma_{\alpha\beta},\qquad\Sigma =\left\Vert\Sigma_{\alpha\beta}\right
\Vert ,\qquad\alpha ,\beta =1,2,\ldots 2N.\label{2.4}\end{equation}

The operator equation (\ref{2.1}) can be transformed into a partial
differential equation upon the choice of some concrete
representation for the statistical operator.  For instance, in the
coordinate representation we write ${\bf q}=({\bf p},{\bf x})$, 
{\bf p} and {\bf x} being the
$N$-vectors, and split the matrix {\bf B} into $N\times N$ blocks:

\[{\bf B}=\left\Vert\begin{array}{cc}
{\bf b}_1&{\bf b}_2\\
{\bf b}_3&{\bf b}_4\end{array}
\right\Vert ,\qquad {\bf b}_1=\widetilde {{\bf b}}_1,\qquad {\bf b}_2
=\widetilde {{\bf b}}_3,\qquad {\bf b}_4=\widetilde {{\bf b}}_4,\qquad {\bf b}_
j=\left\Vert b_j^{mn}\right\Vert\]
(the tilde designates a transposed matrix). Then we obtain the
following second-order equation for the density matrix $\rho (x,x')$:

\begin{eqnarray}
i\hbar\frac {\partial\rho}{\partial t}&=&\frac {\hbar^2}2b_1^{mn}\left
(\frac {\partial^2\rho}{\partial x_m'\partial x_n'}-\frac {\partial^
2\rho}{\partial x_m\partial x_n}\right)-i\hbar b_2^{mn}\left(x_n\frac {
\partial\rho}{\partial x_m}+x_n'\frac {\partial\rho}{\partial x_m'}\right
)\nonumber\\
&+&\frac 12b_4^{mn}\left(x_mx_n-x_m'x_n'\right)\rho -i\hbar\rho\mbox{Tr}
{\bf b}_2\nonumber\\
&-&i\hbar c_1^m\left(\frac {\partial\rho}{\partial x_m}+\frac {\partial
\rho}{\partial x_m'}\right)+c_2^m\left(x_m-x_m'\right).
\label{2.6}\end{eqnarray}
Here ${\bf c}_1$ and ${\bf c}_2$ are $N$-dimensional components of the vector
${\bf C}=({\bf c}_1,{\bf c}_2)$. 
The disadvantage of Eq. (\ref{2.6}) is the broken
symmetry between the coordinates and momenta, which is inherent
in the Hamiltonian. This symmetry can be restored if one proceeds
from the complex density matrix to the real Wigner function

\begin{equation}W({\bf p},{\bf x})=\int\rho ({\bf x}+\xi /2,{\bf x}
-\xi /2)\exp(-i{\bf p}\xi /\hbar )\,{\rm d}\xi ,\label{2.7}\end{equation}
\begin{equation}\rho ({\bf x},{\bf x}')=\int W\left({\bf p},\frac
12({\bf x}+{\bf x}')\right)\exp\left[i{\bf p}({\bf x}-{\bf x}')/\hbar\right
]\mbox{d}{\bf p}/(2\pi\hbar )^N.\label{2.8}\end{equation}
Applying transformation (\ref{2.7}) to Eq.  (\ref{2.6}) we obtain an
equivalent equation that is {\em first order with respect to all
the derivatives\/}:

\begin{equation}\frac {\partial W}{\partial t}=\left(p_nb_2^{nm}+
x_nb_4^{nm}+c_2^m\right)\frac {\partial W}{\partial p_m}-\left(p_
nb_1^{nm}+x_nb_3^{nm}+c_1^m\right)\frac {\partial W}{\partial x_m}
.\label{2.9}\end{equation}

This equation demonstrates the distinction and the advantage of
the Wigner function for the description of quadratic quantum
systems (other remarkable features of the Wigner function were
discussed, e.g., in \cite{Wig-adv}). Eq. (\ref{2.9}) assumes an especially
compact form in terms of the $2N$-vector {\bf q},
\begin{equation}\frac {\partial W}{\partial t}=\frac {\partial}{\partial
q_{\alpha}}\left[(\Sigma {\bf B}{\bf q}+\Sigma {\bf C})_{\alpha}W\right
].\label{2.10}\end{equation}
Here the $2N\times 2N$ matrix $\Sigma$, in accordance with Eq. (\ref{2.4}),
equals
\begin{equation}\Sigma =\left\Vert\begin{array}{cc}
{\bf 0}&{\bf I}_N\\
-{\bf I}_N&{\bf 0}\end{array}
\right\Vert ,\label{2.11}\end{equation}
${\bf I}_N$ being the $N\times N$ unit matrix.

Now let us suppose that we have a {\em given linear\/} equation for the
first-order average values

\begin{equation}\langle\dot {{\bf q}}\rangle ={\bf A}(t)\langle {\bf q}
\rangle +{\bf K}(t),\label{2.12}\end{equation}
with an {\em arbitrary\/} matrix ${\bf A}(t)$ and an 
{\em arbitrary\/} vector ${\bf K}(t)$. The
problem investigated in this section is whether it is possible to
find an equation for the statistical operator or the Wigner
function that would result in the given Eq. (\ref{2.12}). It is
trivial to check that Eq. (\ref{2.12}) is the consequence of the
equation

\begin{equation}\frac {\partial W}{\partial t}=-\frac {\partial}{
\partial q_{\alpha}}\left[({\bf A}{\bf q}+{\bf K})_{\alpha}W\right
].\label{2.14}\end{equation}
However, although Eq. (\ref{2.14}) preserves the normalization of
the Wigner function,
\begin{equation}\int W({\bf q})\,\mbox{d}{\bf q}/(2\pi\hbar )^N=1
,\label{2.15}\end{equation}
it does not agree, in general, with the fundamental quantum
mechanical principle of the positive definiteness of the statistical
operator and (this is almost the same) with the uncertainty
relations. Indeed, let us consider the operator $\widehat {F}=\alpha_
j\left(\widehat q_j-\langle q_j\rangle\right)$
with arbitrary complex coefficients $\alpha_j$. For any stattistical
operator, due to its nonnegative definiteness, the inequality
$\langle\widehat {F}^{\dagger}\widehat {F}\rangle\equiv\mbox{Tr}\left(\widehat
\rho\widehat F^{\dagger}\widehat F\right)\ge 0$ must hold. 
Taking into account the structure
of the operator $\widehat {F}$,
we arrive at the conclusion on the nonnegative
definiteness of the bilinear Hermitian form
\[\alpha^{*}\Phi\alpha\equiv\alpha^{*}_j\Phi_{jk}\alpha_k,\qquad\Phi_{
jk}=\left\langle\left(\widehat q_j-\langle q_j\rangle\right)\left(\widehat
q_k-\langle q_k\rangle\right)\right\rangle ,\]
whose matrix $\Phi$ is constructed from the centered second-order
moments of the operators $\widehat {q}_j$, $j,k=1,2,\ldots ,2N$. 
It is convenient to
distinguish even and odd parts of the matrix $\Phi$. The symmetric part
consists of the symmetrized second moments (covariances)

\begin{equation}{\cal M}_{ij}={\cal M}_{ji}=\frac 12\left\langle\widehat
q_j\widehat q_k+\widehat q_k\widehat q_j\right\rangle -\langle q_j\rangle\langle
q_k\rangle .\label{2.16}\end{equation}
The antisymmetric part, in accordance with Eq. (\ref{2.4}), is
expressed through the commutator matrix, so that
\[\Phi ={\cal M}-\frac {i\hbar}2\Sigma .\]
The elements of the matrix ${\cal M}$ are calculated in terms of the Wigner
function as follows:

\[{\cal M}_{ij}=\int q_iq_jW({\bf q})\,\mbox{d}{\bf q}/(2\pi\hbar
)^N-\bar {q}_i\bar {q}_j,\]
\[\bar {{\bf q}}=\langle {\bf q}\rangle =\int {\bf q}W({\bf q})\,\mbox{d}
{\bf q}/(2\pi\hbar )^N.\]
Due to Eq. (\ref{2.14}), the variance matrix satisfies the equation

\begin{equation}\dot {{\cal M}}={\bf A}{\cal M}+{\cal M}\widetilde {{\bf A}}
.\label{2.18}\end{equation}
The first-order averages do not influence the variances in the case
under study.

If the coefficients $\alpha_j$ are chosen in such a way that the commutator
$\left[\widehat F,\widehat F^{\dagger}\right]$ is positive
(the simplest example is $\widehat {F}=\widehat {x}+i\widehat {p}$), then the
operator $\widehat {F}$ coincides within a constant factor with a boson
annihilation operator. Choosing the initial state to be the vacuum
state for this operator, at the initial instant we have
\[\mbox{Tr}\left(\widehat F^{\dagger}\widehat F\widehat\rho (0)\right)=\alpha^{
*}\Phi (0)\alpha =0.\]
Moreover, the matrix equalities
$\Phi (0)\alpha =\alpha^{*}\Phi (0)=0$ hold as well.
For the chosen initial state we have at $t>0$, due to  Eq.
(\ref{2.18}),

\begin{eqnarray}
\alpha^{*}\Phi (t)\alpha&=&\alpha^{*}\left[{\cal M}(t)-\frac {i\hbar}
2\Sigma\right]\alpha\nonumber\\
&=&\alpha^{*}\left[{\bf A}(0){\cal M}(0)+{\cal M}(0)\widetilde {\bf A}
(0)\right]t\alpha +{\cal O}(t^2)\nonumber\\
&=&\frac 12i\hbar t\alpha^{*}\left[{\bf A}(0)\Sigma +\Sigma\widetilde
{\bf A}(0)\right]\alpha +{\cal O}(t^2).\label{2.19}\end{eqnarray}
For the Hamiltonian systems we have ${\bf A}=-\Sigma {\bf B},\quad
\widetilde {{\bf A}}={\bf B}\Sigma$, and the
linear with respect to time term disappears. But for an arbitrary
matrix ${\bf A}$ the right-hand side of Eq. (\ref{2.19}) can be negative. For
example, in the model of a damped oscillator with the equations
for the averages
\begin{equation}\dot {x}=p,\qquad\dot {p}=-\omega_0^2x-2\gamma p,
\label{2.20}\end{equation}
the matrices entering Eq. (\ref{2.19}) read

\begin{equation}{\bf A}=\left\Vert\begin{array}{cc}
-2\gamma&-\omega_0^2\\
1&0\end{array}
\right\Vert ,\qquad {\bf A}\Sigma +\Sigma\widetilde {{\bf A}}
=\left\Vert\begin{array}{cc}
0&-2\gamma\\
2\gamma&0\end{array}
\right\Vert .\label{2.21}\end{equation}
Then, for the initial vacuum state of operator $\widehat {F}=\widehat {x}+
i\widehat {p}$, when
$\alpha =(i,1)$, we obtain from Eq. (\ref{2.19})
\[\alpha^{*}\Phi (t)\alpha =-2\hbar t\gamma +{\cal O}(t^2)<0.\]

Consequently, in the general case Eq.  (\ref{2.14}) is unacceptable.
Thus we need more complicated generalizations.  The simplest
possibility is to add terms with second derivatives to the right-hand side
of Eq. (\ref{2.14}), i.e., to transform this equation
into the Fokker-Planck equation:

\begin{equation}\frac {\partial W}{\partial t}=-\frac {\partial}{
\partial q_{\alpha}}\left[({\bf A}{\bf q}+{\bf K})_{\alpha}W\right
]+D_{\alpha\beta}\frac {\partial^2W}{\partial q_{\alpha}\partial
q_{\beta}},\label{2.22}\end{equation}
where the diffusion coefficients $D_{\alpha\beta}=D_{\beta\alpha}$, 
combined into a
symmetric matrix ${\bf D}=\left\Vert D_{\alpha\beta}\right\Vert$, 
may depend on time but do not
depend on the coordinates. The new ``diffusion'' terms do not
change the equation for the average values (\ref{2.12}), moreover,
they preserve the normalization (\ref{2.15}). But they enable one to
``save'' the nonnegative definiteness of the statistical operator.
Indeed, considering the evolution of the bilinear form $\alpha^{*}
\Phi (t)\alpha$, we obtain, instead of  Eq. (\ref{2.19}), the equation

\[\alpha^{*}\Phi (t)\alpha =t\alpha^{*}\left(2{\bf D}+\frac {i\hbar}
2\left[{\bf A}(0)\Sigma +\Sigma\widetilde {\bf A}(0)\right]\right)\alpha
+{\cal O}(t^2),\]
since Eq. (\ref{2.18}) is replaced by

\begin{equation}\dot {{\cal M}}={\bf A}{\cal M}+{\cal M}\widetilde {{\bf A}}
+2{\bf D}.\label{2.23}\end{equation}
Consequently, the necessary condition of the compatibility of  Eq.
(\ref{2.22}) with the principles of quantum mechanics is the
nonnegative definiteness of the matrix
\begin{equation}{\bf D}_{*}\equiv {\bf D}+\frac {i\hbar}4\left[{\bf A}
\Sigma +\Sigma\widetilde {\bf A}\right]\ge 0\label{2.24}\end{equation}
{\em at any instant of time}. Moreover, it can be proved
\cite{30,31} that the condition ${\bf D}_{*}\ge 0$ is sufficient as well.

Vector {\bf q} was defined above as ${\bf q}=({\bf p},{\bf x})$, and the
matrix $\Sigma$ had the
explicit form given by Eq. (\ref{2.11}). Let us make the time-independent
transformation of the variables
\[{\bf q}'={\bf T}{\bf q},\qquad\det {\bf T}\neq 0,\qquad\mbox{Im}
{\bf T}\neq 0.\]
Then Eqs.  (\ref{2.12}) and (\ref{2.22}) preserve their forms,
provided the matrices {\bf A} and {\bf D} are replaced by the
matrices ${\bf A}'={\bf T}{\bf A}{\bf T}^{-1}$ and
${\bf D}'={\bf T}{\bf D}\widetilde {{\bf T}}$.  
The nonnegative definiteness of the matrix $
{\bf D}_{*}$ of Eq.
(\ref{2.24}) is equivalent to the nonnegative definiteness of the matrix
${\bf D}_{*}'={\bf T}{\bf D}_{*}\widetilde {{\bf T}}$.  The latter has, 
in turn, again the form (\ref{2.24}), if
one replaces the matrix $\Sigma$ by
$\Sigma'={\bf T}\Sigma\widetilde {{\bf T}}$,
but this is just the
transformation law of any matrix defined according to Eq.
(\ref{2.4}).  This way we arrive at the important conclusion, that
all the formulas containing the matrices {\bf A}, {\bf D},
and $\Sigma$ are valid not only
in the case where the components of the vector {\bf q} coincide with the
canonically conjugate momenta and Cartesian coordinates, but also
in the general case where the components of the vector ${\bf q}$ are
arbitrary Hermitian operators with $c$-number commutators,
provided the matrix $\Sigma$ is defined according to Eq. (\ref{2.4}).

To transform Eq.  (\ref{2.22}) to an operator form that is
independent on the concrete representation, one should take into
account the following correspondence relations between the
operators $\widehat {{\bf q}}\widehat{\rho}$, $\widehat{\rho}\widehat {{\bf q}}$, 
and their Weyl symbols (they result from Eqs.
(\ref{2.7}), (\ref{2.8})):

\begin{equation}\widehat {{\bf q}}\widehat{\rho}\leftrightarrow\left({\bf q}
-\frac {i\hbar}2\Sigma\frac {\partial}{\partial {\bf q}}\right)W(
{\bf q}),\qquad\widehat{\rho}\widehat {{\bf q}}\leftrightarrow\left({\bf q}
+\frac {i\hbar}2\Sigma\frac {\partial}{\partial {\bf q}}\right)W(
{\bf q}),\label{2.25}\end{equation}
\begin{equation}{\bf q}W({\bf q})\leftrightarrow\frac 12\left(\widehat
{\bf q}\widehat\rho +\widehat\rho\widehat {\bf q}\right),\qquad\frac {\partial
W}{\partial {\bf q}}\leftrightarrow\frac i{\hbar}\Sigma^{-1}\left
(\widehat {\bf q}\widehat\rho -\widehat\rho\widehat {\bf q}\right).\label{2.26}\end{equation}
Making transformations (\ref{2.26}) in Eq. (\ref{2.22}) we arrive at
the equation

\begin{eqnarray}
\frac {\partial\widehat{\rho}}{\partial t}&=&\frac i{2\hbar}\left[\widehat
{\bf q}\Sigma^{-1}{\bf A}\widehat {\bf q}\widehat\rho +\widehat\rho\widehat {\bf q}
\widetilde {\bf A}\Sigma^{-1}\widehat {\bf q}+\widehat {\bf q}\left(\Sigma^{-
1}{\bf A}+\widetilde {\bf A}\Sigma^{-1}\right)\widehat\rho\widehat {\bf q}\right
]\nonumber\\
&+&\frac i{\hbar}\left[\widehat {\bf q}\Sigma^{-1}{\bf K}\widehat\rho -\widehat
\rho\widehat {\bf q}\Sigma^{-1}{\bf K}\right]-\frac 1{\hbar^2}\left[\widehat
{\bf q}{\bf S}\widehat {\bf q}\widehat\rho +\widehat\rho\widehat {\bf q}{\bf S}\widehat
{\bf q}-2\widehat {\bf q}{\bf S}\widehat\rho\widehat {\bf q}\right],
\label{2.27}\end{eqnarray}
where the symmetrical matrix
\begin{equation}{\bf S}=-\Sigma^{-1}{\bf D}\Sigma\label{2.28}\end{equation}
must satisfy a constraint equivalent to Eq. (\ref{2.24}):
\begin{equation}{\bf S}_{*}\equiv {\bf S}-\frac {i\hbar}4\left[\widetilde
{\bf A}\Sigma^{-1}+\Sigma^{-1}{\bf A}\right]\ge 0.\label{2.29}\end{equation}
Eqs. (\ref{2.25})-(\ref{2.29}) hold for any vector {\bf q} with a $c$-number
commutator matrix $\Sigma$ of (\ref{2.4}).

Comparison of the elegant equation (\ref{2.22}) for the Wigner
function with the much more cumbersome equation (\ref{2.27})
demonstrates once more the advantage of the Wigner
representation for describing quantum systems with linear
equations of motion for the averages.

There is an important difference between the quantum
Fokker-Planck equation (\ref{2.22}) and its classical counterpart.
The classical Fokker-Planck equation contains, as a rule, second
derivatives only with respect to momenta. However, such a simple
set of the diffusion coefficients is unacceptable in the quantum
case (although sometimes this incorrect equation was considered:  see,
e.g., \cite{Walls-Milb}).
This statement can be easily demonstrated on the example of
system (\ref{2.20}). Writing the matrix {\bf D} in the form
\[{\bf D}=\left\Vert\begin{array}{cc}
D_p&D_{px}\\
D_{px}&D_x\end{array}
\right\Vert\]
and taking into account Eq. (\ref{2.21}) we obtain the matrix
\[{\bf D}_{*}=\left\Vert\begin{array}{cc}
D_p&D_{px}-i\hbar\gamma /2\\
D_{px}+i\hbar\gamma /2&D_x\end{array}
\right\Vert .\]
The condition of its positive definiteness is given by the inequality
\cite{32}-\cite{35}
\begin{equation}\det {\bf D}_{*}\equiv D_pD_x-D_{px}^2-\hbar^2\gamma^
2/4\ge 0,\label{2.37}\end{equation}
whose violation leads to the violation of the uncertainty relations
\cite{32}, \cite{36}.  For an arbitrary one-dimensional system
(\ref{2.12}) the condition ${\bf D}_{*}\ge 0$ is equivalent to the inequality
\cite{31}
\begin{equation}\det {\bf D}\ge\hbar^2(\mbox{Tr}{\bf A})^2/16.
\label{2.38}\end{equation}

\setcounter{equation}{0}
\section{Fokker-Planck equation for a subsystem}

In the preceding section we have shown that any given equation
(\ref{2.12}) {\em may be\/} considered as a consequence of some suitable
Fokker-Planck equation for the Wigner function.  The only problem
is to select the set of diffusion coefficients satisfying the
condition ${\bf D}_{*}\ge 0$.  It is clear that by taking sufficiently
large diffusion
coefficients one can always satisfy this condition.  The problem of
finding ``minimal admissible'' diffusion coefficients for some simple
systems (such as a one-dimensional harmonic oscillator or a
two-dimensional isotropic oscillator in a uniform magnetic field)
was investigated in \cite{30},\cite{37},\cite{38}.

Here we investigate the following problem.  Suppose we have a
large {\em closed\/} quantum system with $N$ degrees of freedom, described
by Hamiltonian (\ref{1.1}).  Let us split the
vector ${\bf q}$ in two parts:
${\bf q}=({\bf Q},\xi )$, where the $n$-dimensional vector ${\bf Q}$
describes a {\em subsystem}, while the vector $\xi$ relates to
a {\em reservoir}.  The question is:  what kind
of equation describes the evolution of the {\em subsystem\/} if one
performs an averaging over the variables of the reservoir?

The Wigner function of the whole system is given by the relation

\begin{equation}W({\bf q},t)=\int G({\bf q},{\bf q}',t)W({\bf q}'
,0)\,\mbox{d}{\bf q}',\label{3.1}\end{equation}
where the propagator $G({\bf q},{\bf q}',t)$ satisfies Eq. (\ref{2.10}) and the
initial condition $G({\bf q},{\bf q}',0)=\delta ({\bf q}-{\bf q}')$.
Since Eq. (\ref{2.10}) is
first-order with respect to all the variables, its propagator is
extremely simple:
\begin{equation}G({\bf q},{\bf q}',t)=\delta\left({\bf q}-{\bf q}_{
*}(t;{\bf q}')\right),\label{3.2}\end{equation}
where the vector
\begin{equation}{\bf q}_{*}(t;{\bf q}')={\bf R}(t)\left[{\bf q}'-
\Delta (t)\right]\label{3.3}\end{equation}
is the solution to the classical equation of motion (\ref{2.2}),
satisfying the initial condition ${\bf q}_{*}(0;{\bf q}')={\bf q}'$. 
Consequently, the
$2N\times 2N$ matrix ${\bf R}(t)$ satisfies the equation
\begin{equation}\dot {{\bf R}}=-\Sigma {\bf B}{\bf R}\equiv {\cal A}
{\bf R},\label{3.5}\end{equation}
and the initial condition ${\bf R}(0)={\bf I}_{2N}$. The vector $\Delta
(t)$ equals zero at
$t=0$. For $t>0$ it is determined from the equations
\begin{equation}\dot{\Delta }={\bf R}^{-1}\Sigma {\bf C}\equiv\Sigma
\widetilde {{\bf R}}{\bf C}.\label{3.6}\end{equation}
These two forms are equivalent due to the identities
\begin{equation}\widetilde {{\bf R}}(t)\Sigma^{-1}{\bf R}(t)\equiv\Sigma^{
-1},\qquad {\bf R}(t)\Sigma\widetilde {{\bf R}}(t)\equiv\Sigma ,\qquad
{\bf R}^{-1}\equiv\Sigma\widetilde {{\bf R}}\Sigma^{-1},
\label{3.7}\end{equation}
which follow from Eq. (\ref{3.5}).

It is clear that the Heisenberg equation of motion for the operator
$\widehat {{\bf q}}$
coincides with Eq. (\ref{2.2}), provided $\langle {\bf q}\rangle$
is replaced by $\widehat {{\bf q}}$. The
solution to this equation is given by Eq. (\ref{3.3}) with carets
over ${\bf q}$ and ${\bf q}'$. Therefore, the identities (\ref{3.7})
mean nothing but the
conservation of the commutation relations (\ref{2.4}) in time, or, in
other words, the canonicity of transformation (\ref{3.3}) and the
unitarity of the evolution operator.

Now let us proceed to the averaging of the total Wigner function over
the reservoir variables $\xi$. Our first assumption is that the initial
total Wigner function is factorized:
\begin{equation}W({\bf q},0)=W_0({\bf q})W_1(\xi ).\label{3.12}\end{equation}
The second assumption concerns the initial Wigner function of the
reservoir. We assume it to be {\em Gaussian},
\begin{equation}W_1(\xi )=\hbar^M(\det {\bf F})^{-1/2}\exp\left[-\frac
12(\xi -\gamma ){\bf F}^{-1}(\xi -\gamma )\right],\label{3.13}\end{equation}
with some symmetric positive definite $2M\times 2M$ matrix {\bf F} and a
$2M$-vector $\gamma$ ($M$ being the number of degrees of freedom of the
reservoir).  In particular, it may correspond to a mixed equilibrium
state \cite{8,15,Akh-82} or to a pure squeezed coherent
state \cite{7,15,Schum}.  For any physically admissible
Gaussian Wigner function the covariance matrix {\bf F} must satisfy a
set of conditions expressing generalized uncertainty relations.
The simplest among them is the inequality \cite{30,39}
$\det {\bf F}\ge (\hbar^2/4)^M$. Moreover, the parameter
\begin{equation}\mu =(\hbar /2)^M(\det {\bf F})^{-1/2}\le 1
\label{3.14}\end{equation}
characterizes ``the degree of quantum mechanical purity'' of the
Gaussian state:

\begin{equation}\mu =\mbox{Tr}\widehat{\rho}^2=\int W^2(\xi )\,\mbox{d}\xi
/(2\pi\hbar )^M.\label{3.15}\end{equation}

To calculate the averaged Wigner function
\begin{equation}W_{\xi}({\bf Q},t)=\int W({\bf Q},\xi )\mbox{d}\xi
/(2\pi\hbar )^{^M}\label{2.13}\end{equation}
we split the matrix {\bf R} and the vector $\Delta$ into rectangular blocks in
accordance with the decomposition ${\bf q}=({\bf Q},\xi )$:
\begin{equation}{\bf R}=\left\Vert\begin{array}{cc}
{\bf R}_{11}&{\bf R}_{12}\\
{\bf R}_{21}&{\bf R}_{22}\end{array}
\right\Vert ,\qquad\Delta =\left\Vert\begin{array}{c}
\Delta_{{\bf Q}}\\
\Delta_{\xi}\end{array}
\right\Vert .\label{3.17}\end{equation}
Then Eqs. (\ref{3.1}), (\ref{3.2}), (\ref{3.12}) lead to the integral
\begin{eqnarray}
W_{\xi}({\bf Q},t)&=&\int\delta\left({\bf Q}-{\bf R}_{11}\left[{\bf Q}'
-\Delta_{{\bf Q}}\right]-{\bf R}_{12}\left[\xi'-\Delta_{\xi}\right
]\right)\nonumber\\
&&\times\delta\left(\xi -{\bf R}_{21}\left[{\bf Q}'-\Delta_{{\bf Q}}\right
]-{\bf R}_{22}\left[\xi'-\Delta_{\xi}\right]\right)\nonumber\\
&&\times W_0({\bf Q}')W_1(\xi')\mbox{d}{\bf Q}'\mbox{d}\xi'\mbox{d}
\xi /(2\pi\hbar )^M\label{3.16}\end{eqnarray}
The integration over $\mbox{d}\xi$ is trivial: it simply removes the second
delta function. To perform the integration over $\mbox{d}\xi'$ we replace the
first delta function with its integral representation (recall that
the dimension of the vector {\bf Q} is $2n$),
\begin{equation}\delta ({\bf x})=\int e^{i{\bf k}{\bf x}}\,\mbox{d}^{
2n}{\bf k}/(2\pi )^{2n}.\label{3.18}\end{equation}
Thus we obtain two {\em Gaussian\/} integrals (the first one over $\mbox{d}
\xi'$ and
the second over $\mbox{d}{\bf k}$) that can be calculated exactly due to the
well-known formula
\begin{equation}\int_{-\infty}^{\infty}\mbox{d}{\bf x}\exp(-{\bf x}
{\bf A}{\bf x}+{\bf b}{\bf x})=[\det({\bf A}/\pi )]^{-1/2}\exp\left
(\frac 14{\bf b}{\bf A}^{-1}{\bf b}\right).\label{3.19}\end{equation}
Finally we arrive at the same equation (\ref{3.1}), but with the
variables ${\bf Q},{\bf Q}'$ instead of ${\bf q},{\bf q}'$, $W_{\xi}
({\bf Q},t)$ instead of $W({\bf q},t)$, and with
the {\em averaged propagator}
\begin{eqnarray}
G_{\xi}({\bf Q},{\bf Q}',t)&=&(2\pi )^{-n}\left[\det {\cal M}_{*}
(t)\right]^{-1/2}\nonumber\\
&\times&\exp\left[-\frac 12\left({\bf Q}-{\bf R}_{11}{\bf Q}'-\delta_{
*}\right){\cal M}_{*}^{-1}\left({\bf Q}-{\bf R}_{11}{\bf Q}'-\delta_{
*}\right)\right],\label{3.20}\end{eqnarray}
where the symmetric matrix ${\cal M}_{*}(t)$ equals
\begin{equation}{\cal M}_{*}(t)={\bf R}_{12}(t){\bf F}\widetilde {{\bf R}}_{
12}(t).\label{3.21}\end{equation}
Recall that the dimension of the rectangular matrix ${\bf R}_{12}$ is
$2n\times 2M$. The vector $\delta_{*}(t)$ equals
\begin{equation}\delta_{*}(t)={\bf R}_{12}(t)\gamma -{\bf R}_{11}
(t)\Delta_{{\bf Q}}(t)-{\bf R}_{12}(t)\Delta_{\xi}(t).\label{3.22}\end{equation}

Direct inspection shows that the propagator (\ref{3.20})
(and, consequently, its convolution with any initial function) satisfies
the Fokker-Planck equation (\ref{2.22}) with the following drift
matrix ${\bf A}$ and vector ${\bf K}$,
\begin{equation}{\bf A}=\dot {{\bf R}}_{11}{\bf R}_{11}^{-1},
\label{3.25}\end{equation}
\begin{equation}{\bf K}=\dot{\delta}_{*}-{\bf A}\delta_{*}.
\label{3.26}\end{equation}
The matrix of diffusion coefficients reads
\begin{equation}{\bf D}=\frac 12\left(\dot {\cal M}_{*}-{\bf A}{\cal M}_{
*}-{\cal M}_{*}\widetilde {\bf A}\right)=\mbox{sym}\left[\left(\dot {\bf R}_{
12}-\dot {\bf R}_{11}{\bf R}_{11}^{-1}{\bf R}_{12}\right){\bf F}\widetilde
{\bf R}_{12}\right],\label{3.27}\end{equation}
where we have introduced the notation
\begin{equation}\mbox{sym}{\bf A}\equiv\frac 12\left({\bf A}+\widetilde
{\bf A}\right).\label{5.14}\end{equation}

Thus we have proved that the evolution of the Wigner function of
any subsystem of a closed {\em quadratic\/} quantum system is governed
by some effective Fokker-Planck equation, {\em provided the initial
state of the ``remaining part of the system'' is Gaussian}. However, the
coefficients of the equation  obtained in this way depend, as a rule, on
time even for a time-independent Hamiltonian of the closed system.

\setcounter{equation}{0}
\section{Two coupled oscillators}

Let us illustrate the formulas of the preceding section,
applying them to a very simple special model, where both the
``subsystem under study'' and the ``reservoir'' are the harmonic
oscillators with a single degree of freedom.  This example admits
{\em exact\/} solutions, in contrast to the more realistic situation of
a reservoir with a very large number of degrees of freedom, where
one has to make various simplifications and approximations to
obtain a closed result that would be easy to analyze.  Consider
first the free Hamiltonian
\begin{equation}H_0=\frac {p_1^2}{2m_1}+\frac 12m_1\omega_1^2x_1^
2+\frac {p_2^2}{2m_2}+\frac 12m_2\omega_2^2x_2^2\label{4.0}\end{equation}
and the most general quadratic interaction Hamiltonian
\begin{equation}H_{\mbox{int}}=g_{pp}p_1p_2+g_{px}p_1x_2+g_{xp}x_
1p_2+g_{xx}x_1x_2.\label{4.1}\end{equation}
We assume all the coefficients to be time-independent.  In the notation
of the preceding section we should write
\[{\bf q}=\left(p_1,x_1,p_2,x_2\right),\qquad {\bf Q}=\left(p_1,x_
1\right),\qquad\xi =\left(p_2,x_2\right).\]
Then the matrix ${\cal A}=-\Sigma {\bf B}$ (see Eq.  (\ref{3.5})) reads
\begin{equation}{\cal A}=\left\Vert\begin{array}{cccc}
0&-m_1\omega_1^2&-g_{xp}&-g_{xx}\\
m_1^{-1}&0&g_{pp}&g_{px}\\
-g_{px}&-g_{xx}&0&-m_2\omega_2^2\\
g_{pp}&g_{xp}&m_2^{-1}&0\end{array}
\right\Vert .\label{4.3}\end{equation}
Its characteristic equation turns out to be biquadratic:
\begin{equation}\det({\cal A}-i\omega {\bf I})=\omega^4-\left(\omega_
1^2+\omega_2^2+2\Delta\right)\omega^2+\omega_1^2\omega_2^2+\Delta^
2-g=0,\label{4.4}\end{equation}
where
\begin{equation}\Delta =\det {\cal A}_{12}=g_{pp}g_{xx}-g_{px}g_{
xp},\label{4.5}\end{equation}
\begin{equation}g=\frac {g_{xx}^2}{m_1m_2}+\frac {m_1}{m_2}\omega_
1^2g_{px}^2+\frac {m_2}{m_1}\omega_2^2g_{xp}^2+m_1m_2\omega_1^2\omega_
2^2g_{pp}^2.\label{4.6}\end{equation}
The solutions of Eq.  (\ref{4.4}) are as follows:
\begin{eqnarray*}
\omega_{\pm}&=&\frac 1{\sqrt {2}}\left(\left[\frac 12\left(\omega_
1^2+\omega_2^2\right)+\left(\omega_1^2\omega_2^2+\Delta^2-g\right
)^{1/2}+\Delta\right]^{\frac 12}\right.\nonumber\\
&&\left.\pm\left[\frac 12\left(\omega_1^2+\omega_2^2\right)-\left
(\omega_1^2\omega_2^2+\Delta^2-g\right)^{1/2}+\Delta\right]^{\frac
12}\right)\label{4.7}\end{eqnarray*}
(two other solutions are equal to $-\omega_{\pm}$).

Thus, the evolution of two coupled harmonic oscillators can be
described explicitly for quite arbitrary quadratic interaction
Hamiltonians with time-independent coefficients.  If both
frequences given by Eq.  (\ref{4.7}) are real, then the
particles perform harmonic oscillations.  For certain
parameters complex normal frequencies are possible.  Then the
motion becomes aperiodic.  However, since any normal frequency
$\omega_{+}$ or $\omega_{-}$ is accompanied by the frequency
with the opposite sign,
it is impossible to obtain {\em damped\/} oscillations of either
particle.  The coordinate and momentum of any oscillator
will increase with time.

To illustrate this statement, let us consider first the case (which
seems the most natural) of the interaction via the coordinates,
where the only nonzero coefficient in Eq.  (\ref{4.1}) is $g_{xx}$.  Then
$\Delta =0$, and for sufficiently strong coupling, when
$g>\omega_1^2\omega_2^2$, we have the real frequency $\omega_{+}$
and the pure imaginary frequency $\omega_{-}$:
\begin{equation}\omega_{+}\equiv\omega =\left\{\left[g+\frac 14\left
(\omega_1^2-\omega_2^2\right)^2\right]^{1/2}+\frac 12\left(\omega_
1^2+\omega_2^2\right)\right\}^{\frac 12},\label{4.8}\end{equation}
\begin{equation}\omega_{-}\equiv i\lambda ,\qquad\lambda =\left\{\left
[g+\frac 14\left(\omega_1^2-\omega_2^2\right)^2\right]^{1/2}-\frac
12\left(\omega_1^2+\omega_2^2\right)\right\}^{\frac 12}.
\label{4.9}\end{equation}

Solving Eq. (\ref{3.5}), we obtain the following formula for the matrix
${\bf R}_{11}(t)$:
\begin{equation}{\bf R}_{11}=\left\Vert\begin{array}{cccc}
\rho_{+}\cos\omega t+\rho_{-}\cosh\lambda t&&m_1\left(-\omega\rho_{
+}\sin\omega t+\lambda\rho_{-}\sinh\lambda t\right)\\
&&\\
\frac 1{m_1}\left(\frac {\rho_{+}}{\omega}\sin\omega t+\frac {\rho_{
-}}{\lambda}\sinh\lambda t\right)&&\rho_{+}\cos\omega t+\rho_{-}\cosh
\lambda t&\end{array}
\right\Vert ,\label{4.10}\end{equation}
where
\begin{equation}\rho_{\pm}=\frac 12\pm\frac 14\left(\omega_1^2-\omega_
2^2\right)\left[g+\frac 14\left(\omega_1^2-\omega_2^2\right)^2\right
]^{-\frac 12}.\label{4.11}\end{equation}
We see that the oscillations of the first particle actually increase.
Its energy is derived from the nonpositive definite potential
energy of the whole system, since the whole system turns out to be
{\em unstable\/} when $g>\omega_1^2\omega_2^2$.  Note that Eq.
(\ref{4.10}) still holds
even if $g_{xp}\ne 0$.  If $g_{xx}=g_{xp}=0$, but $g_{pp}\ne 0$ and $
g_{px}\ne 0$,
we obtain, instead of Eq. (\ref{4.10}), the expression
\begin{equation}{\bf R}_{11}=\left\Vert\begin{array}{cccc}
\rho_{+}\cos\omega t+\rho_{-}\cosh\lambda t&-m_1\omega_1^2\left(\frac {
\rho_{+}}{\omega}\sin\omega t+\frac {\rho_{-}}{\lambda}\sinh\lambda
t\right)\\
&\\
\frac 1{m_1\omega_1^2}\left(\omega\rho_{+}\sin\omega t+\lambda\rho_{
-}\sinh\lambda t\right)&\rho_{+}\cos\omega t+\rho_{-}\cosh\lambda
t&\end{array}
\right\Vert ,\label{4.12}\end{equation}
with the same values of $\omega$ and $\lambda$ (provided, of course, $
g>\omega_1^2\omega_2^2$).

Both frequencies $\omega_{+}$ and $\omega_{-}$ have nonzero imaginary parts
provided the argument of the first square bracket in Eq. (\ref{4.7})
is positive while the argument of the second square bracket is
negative.  Then the following inequality must hold:
\begin{equation}\frac 14\left(\omega_1^2-\omega_2^2\right)^{^2}+\Delta\left
(\omega_1^2+\omega_2^2\right)+g<0.\label{4.13}\end{equation}
Assuming for the sake of simplicity that $\omega_1=\omega_2=\omega_0$,
we can rewrite it as
\begin{equation}\left(m_1m_2\right)^{-1}\left[\left(m_1g_{xx}+m_2
\omega_0^2g_{pp}\right)^2+\omega_0^2\left(m_1g_{px}-m_2g_{xp}\right
)^2\right]<0.\label{4.14}\end{equation}
If $\omega_0^2>0$, the only possibility for fulfilling this inequality
is to assume
that the masses have opposite signs.  Of course, such a system is
unstable, and it can be considered only as an extremely simplified
model. Nonetheless, sometimes the models of this sort were
considered \cite{Glaub-Zven}. So we discuss briefly this case as
well. Suppose for simplicity that $m_1=1$, $m_2=-1$. Then $g<0$. The square
roots in Eq. (\ref{4.7}) can be extracted if $|g|=2\omega_0^2|\Delta|$.
In this case the imaginary parts of $\omega_{\pm}$ differ from zero,
provided $\Delta <0$. Such a
situation holds for the following relations between the coupling
coefficients:
\begin{equation}g_{xp}=g_{px},\qquad g_{xx}=-\omega_0^2g_{pp}.
\label{4.17}\end{equation}
Then
\begin{equation}\omega_{\pm}=\omega_0\pm i\gamma ,\qquad\gamma =|
\Delta |^{1/2}=\left(g_{px}^2+\omega_0^2g_{pp}^2\right)^{1/2}.
\label{4.19}\end{equation}
The explicit form of the matrix ${\bf R}_{11}$ reads
\begin{equation}{\bf R}_{11}=\cosh\gamma t\left\Vert\begin{array}{ccc}
\cos\omega_0t&\quad&-\omega_0\sin\omega_0t\\
\omega_0^{-1}\sin\omega_0t&\quad&\cos\omega_0t\end{array}
\right\Vert .\label{4.20}\end{equation}
Consequently, the amplitude of the oscillations increases without bound.
The drift matrix ${\bf A}$ (\ref{3.25}) depends on time as follows:
\begin{equation}{\bf A}=\left\Vert\begin{array}{cc}
\gamma\tanh\gamma t&-\omega_0^2\\
1&\gamma\tanh\gamma t\end{array}
\right\Vert .\label{4.21}\end{equation}

An interesting model described by the Hamiltonian
\begin{equation}H=p_1p_2+\omega_0^2x_1x_2+\gamma\left(x_2p_2-x_1p_
1\right)\label{4.22}\end{equation}
was proposed for the first time by Bateman \cite{40}. The
Lagrangian
\begin{equation}L=\dot {x}_1\dot {x}_2-\left(\omega_0^2+\gamma^2\right
)x_1x_2+\gamma\left(x_1\dot x_2-\dot x_1x_2\right)\label{4.24}\end{equation}
was considered by Morse and Feshbach \cite{41}.  This system
was investigated, e.g.,  in \cite{mirror} (see also \cite{Dekker}).
The equations of motion in this case read

\begin{equation}\ddot {x}_1+2\gamma\dot {x}_1+\left(\omega_0^2+\gamma^
2\right)x_1=0,\qquad\ddot {x}_2+2\gamma\dot {x}_2+\left(\omega_0^
2+\gamma^2\right)x_2=0.\label{4.23}\end{equation}
Consequently, {\em at the classical level\/} we have damping in the first
mode and amplification in the second mode.  Moreover, at the
classical level both particles are completely independent due to Eq.
(\ref{4.23}).  The quantum picture is more complicated, since the
quantum behaviour is governed not by the second-order equations
of motion, but by the {\em Hamiltonian\/} (\ref{4.22}), in which the
dynamical variables of both particles are entangled.  (The
nonunique correspondence between the equations of motion and the
Lagrangians or Hamiltonians leading to them, as well as related
ambiguities of quantization, were investigated in detail in
\cite{Skarzh}. It can be proved  \cite{Skarzh,Havas} that no
Hamiltonian leading to Eq. (\ref{4.23}) and coinciding with Eq.
(\ref{4.0}) for $\gamma =0$ exists.) In particular, the velocity of each
particle is determined by the generalized momentum of {\em the other \/}
particle:
\[\dot {x}_1=p_2-\gamma x_1,\qquad\dot {x}_2=p_1+\gamma x_2.\]
The two other Hamilton equations read
\[\dot {p}_1=\gamma p_1-\omega_0^2x_2,\qquad\dot {p}_2=\gamma p_2
-\omega_0^2x_1.\]
In terms of the blocks of the matrix ${\bf R}$ these
equations can be rewritten as follows:
\begin{equation}\dot {{\bf R}}_{1k}={\bf a}{\bf R}_{1k}+{\bf b}{\bf R}_{
2k},\qquad\dot {{\bf R}}_{2k}={\bf b}{\bf R}_{1k}-{\bf a}{\bf R}_{
2k},\qquad (k=1,2)\label{4.26}\end{equation}
where {\bf a} and {\bf b} are $2\times 2$ matrices,
\[{\bf a}=\left\Vert\begin{array}{cc}
\gamma&0\\
0&-\gamma\end{array}
\right\Vert ,\qquad {\bf b}=\left\Vert\begin{array}{cc}
0&-\omega_0^2\\
1&0\end{array}
\right\Vert .\]
Eliminating the matrix ${\bf R}_{2k}$ we obtain the second-order equation
\begin{equation}\ddot {{\bf R}}_{1k}-2{\bf a}\dot {{\bf R}}_{1k}+\left
(\gamma^2+\omega_0^2\right){\bf R}_{1k}=0.\label{4.27}\end{equation}
Seeking its solution in the form ${\bf R}_{1k}=\exp(\Lambda t
){\bf R}_0$ we obtain the
characteristic equation
\[\Lambda^2-2{\bf a}\Lambda +\left(\gamma^2+\omega_0^2\right){\bf I}
=0,\]
whose solution reads
\[\Lambda ={\bf a}\pm i\omega_0{\bf I}.\]
Taking into account the initial conditions
\[{\bf R}_{11}(0)={\bf I},\quad\dot {{\bf R}}_{11}(0)={\bf a},\qquad
{\bf R}_{12}(0)=0,\quad\dot {{\bf R}}_{12}(0)={\bf b},\]
we obtain finally the matrices
\[{\bf R}_{11}=\cos\omega_0t\left\Vert\begin{array}{cc}
e^{\gamma t}&0\\
0&e^{-\gamma t}\end{array}
\right\Vert ,\qquad {\bf R}_{12}=\sin\omega_0t\left\Vert\begin{array}{cc}
0&-\omega_0e^{\gamma t}\\
\omega_0^{-1}e^{-\gamma t}&0\end{array}
\right\Vert .\]
Consequently, after averaging over the state of the second particle
we have a system with a damped coordinate, but with a momentum
increasing in time without bound.  Moreover, the momentum
will no longer be related to the velocity.  The drift matrix
(\ref{3.25}) in the Fokker-Planck equation for the averaged Wigner
function of the first particle depends on time as follows:
\[{\bf A}=\left\Vert\begin{array}{cc}
\omega_0\tan(\omega_0t)+\gamma&0\\
0&\omega_0\tan(\omega_0t)-\gamma\end{array}
\right\Vert .\]

We see that the reduction of ``Bateman's mirror model'' (\ref{4.22})
does not lead to a damped quantum oscillator in the conventional
meaning of this term.

\setcounter{equation}{0}
\section{Oscillator in a thermostat: weak coupling limit}

We now proceed to a more realistic model, where the oscillator
under study (its frequency will be denoted by $\omega_0$) is coupled to a
large number of other oscillators with frequencies $\omega_i$. This model 
was the subject of investigations in many papers: see, e.g.,
\cite{16}-\cite{O'Con}. A more comprehensive reference list can be 
found in the review \cite{Dekker}. We assume that each oscillator is
described by the standard Hamiltonian
\[H_i=\frac 12\left(p_i^2+\omega_i^2x_i^2\right)\]
with unit mass (this can easily be achieved by rescaling
the coordinates), while the quadratic interaction Hamiltonian is 
chosen in the most general form:  
\begin{equation}H_{\mbox{int}}=\sum_i\left(z_ip_ip_0+v_ip_ix_0+u_
ix_ip_0+g_ix_ix_0\right).\label{5.1}\end{equation}
The coupling constants $z_i$, $v_i$, $u_i$, $g_i$ are assumed to be time
independent. 

Equation  (\ref{3.5}) and its initial condition are equivalent to the 
following equations and initial conditions for the blocks of the matrix
${\bf R}$ of (\ref{3.17}):
\begin{equation}\dot {{\bf R}}_{11}={\cal A}_{11}{\bf R}_{11}+{\cal A}_{
12}{\bf R}_{21},\qquad {\bf R}_{11}(0)={\bf I},\label{5.2}\end{equation}
\begin{equation}\dot {{\bf R}}_{21}={\cal A}_{21}{\bf R}_{11}+{\cal A}_{
22}{\bf R}_{21},\qquad {\bf R}_{21}(0)=0,\label{5.3}\end{equation}
\begin{equation}\dot {{\bf R}}_{12}={\cal A}_{11}{\bf R}_{12}+{\cal A}_{
12}{\bf R}_{22},\qquad {\bf R}_{12}(0)=0,\label{5.4}\end{equation}
\begin{equation}\dot {{\bf R}}_{22}={\cal A}_{21}{\bf R}_{12}+{\cal A}_{
22}{\bf R}_{22},\qquad {\bf R}_{22}(0)={\bf I}.\label{5.5}\end{equation}
The $2\times 2$ matrix ${\cal A}_{11}$ and the $2M\times 2M$ matrix
${\cal A}_{22}$ ($M$ is the number of
oscillators in the reservoir) read
\begin{equation}{\cal A}_{11}=\left\Vert\begin{array}{cc}
0&-\omega_0^2\\
1&0\end{array}
\right\Vert ,\qquad\left\Vert\begin{array}{cc}
0&-\mbox{diag}\left(\omega_1^2,\ldots ,\omega_i^2,\ldots\right)\\
\mbox{diag}(1,\ldots ,1,\ldots )&0\end{array}
\right\Vert .\label{5.6}\end{equation}
The matrices ${\cal A}_{12}$ and ${\cal A}_{21}$ are rectangular
with the dimensions $2\times 2M$
and $2M\times 2$, respectively:
\begin{equation}{\cal A}_{12}=\left\Vert\begin{array}{cccccccc}
-v_1&\cdots&-v_i&\cdots&-g_1&\cdots&-g_i&\cdots\\
z_1&\cdots&z_i&\cdots&u_1&\cdots&u_i&\cdots\end{array}
\right\Vert ,\label{5.7}\end{equation}
\begin{equation}{\cal A}_{21}=\left\Vert\begin{array}{cc}
-u_1&-g_1\\
\vdots&\vdots\\
-u_i&-g_i\\
\vdots&\vdots\\
z_1&v_1\\
\vdots&\vdots\\
z_i&v_i\\
\vdots&\vdots\end{array}
\right\Vert .\label{5.8}\end{equation}

In this section we consider the case where all the elements of the
interaction matrices ${\cal A}_{12}$ and ${\cal A}_{21}$ are {\em small}.
Then we may use
perturbation theory.  In the zeroth approximation we obtain from Eq.
(\ref{5.2}) 
\begin{equation}{\bf R}_{11}^{(0)}(t)=\exp\left({\cal A}_{11}t\right
).\label{5.9}\end{equation}
Putting this expression into the right-hand side of Eq. (\ref{5.3}) we
obtain the first-order solution for the matrix ${\bf R}_{21}$,
\begin{equation}{\bf R}_{21}^{(1)}(t)=\exp\left({\cal A}_{22}t\right
)\int_0^t\exp\left(-{\cal A}_{22}\tau\right){\cal A}_{21}\exp\left
({\cal A}_{11}\tau\right)\,\mbox{d}\tau .\label{5.10}\end{equation}
Then Eqs.  (\ref{3.25}) and (\ref{5.2}) lead to the following
first-order approximation for the matrix ${\bf A}$ governing the evolution of
the average values of the subsystem variables according to Eq.
(\ref{2.22}):  
\begin{equation}{\bf A}^{(1)}={\cal A}_{11}+{\cal A}_{12}{\bf R}_{
21}^{(1)}(t)\left[{\bf R}_{11}^{(0)}(t)\right]^{-1}.\label{5.11}\end{equation}

Taking into account Eqs. (\ref{5.9}) and (\ref{5.10}) and making the
change of variable $t-\tau =x$ in the integrand we arrive at the 
formula
\begin{equation}\mu\equiv {\bf A}-{\cal A}_{11}={\cal A}_{12}\int_
0^t\exp\left({\cal A}_{22}x\right){\cal A}_{21}\exp\left(-{\cal A}_{
11}x\right)\,\mbox{d}x.\label{5.12}\end{equation}

The solutions to Eqs. (\ref{5.4}) and (\ref{5.5}) in the same
approximation read
\begin{equation}{\bf R}_{22}^{(0)}(t)=\exp\left({\cal A}_{22}t\right
),\label{5.15}\end{equation}
\begin{equation}{\bf R}_{12}^{(1)}(t)=\exp\left({\cal A}_{11}t\right
)\int_0^t\exp\left(-{\cal A}_{11}\tau\right){\cal A}_{12}\exp\left
({\cal A}_{22}\tau\right)\,\mbox{d}\tau .\label{5.16}\end{equation}
Then the diffusion matrix of the Fokker-Planck equation 
(\ref{2.22}), due to Eqs. (\ref{3.27}) and (\ref{5.4}), reads
\begin{eqnarray}
{\bf D}&=&\mbox{sym}\left({\cal A}_{12}{\bf R}_{22}^{(0)}(t){\bf F}
\tilde {\bf R}_{12}^{(1)}(t)\right)\nonumber\\
&=&\mbox{sym}\left\{{\cal A}_{12}\exp\left({\cal A}_{22}t\right){\bf F}
\int_0^t\exp\left(\tilde {\cal A}_{22}\tau\right)\tilde {\cal A}_{
12}\exp\left(-\tilde {\cal A}_{11}\tau\right)\,\mbox{d}\tau\exp\left
(\tilde {\cal A}_{11}t\right)\right\}.\label{5.17}\end{eqnarray}

The explicit forms of the matrices ${\bf R}_{11}^{(0)}(t)$ and
${\bf R}_{22}^{(0)}(t)$ are as follows:
\begin{equation}\exp\left({\cal A}_{11}t\right)=\left\Vert\begin{array}{cc}
\cos\omega_0t&-\omega_0\sin\omega_0t\\
\omega_0^{-1}\sin\omega_0t&\cos\omega_0t\end{array}
\right\Vert ,\label{5.21}\end{equation}
\begin{equation}\exp\left({\cal A}_{22}t\right)=\left\Vert\begin{array}{cc}
\mbox{diag}(\cos\omega_it)&\mbox{diag}(-\omega_i\sin\omega_it)\\
\mbox{~diag}(\omega_i^{-1}\sin\omega_it)&\mbox{diag}(\cos\omega_i
t)\end{array}
\right\Vert ,\label{5.22}\end{equation}
If we choose the thermostat variance matrix to be
\begin{equation}{\bf F}=\left\Vert\begin{array}{cc}
\mbox{diag}\left(\omega_i^2f_i\right)&0\\
0&\mbox{diag}\left(f_i\right)\end{array}
\right\Vert ,\label{5.30}\end{equation}
(in particular, ${\bf F}$ may be an equilibrium variance matrix for the 
thermostat variables), then it describes the steady-state solution 
of Eq.  (\ref{2.18}) in the absense of the interaction:
\begin{equation}{\bf F}(t)=\exp\left({\cal A}_{22}t\right){\bf F}\exp\left
(\tilde {\cal A}_{22}t\right)={\bf F}.\label{5.18}\end{equation}
Consequently, one may interchange the matrices
$\exp\left({\cal A}_{22}t\right)$ and ${\bf F}$ in Eq.
(\ref{5.17}) in accordance with the formula
\begin{equation}\exp\left({\cal A}_{22}t\right){\bf F}={\bf F}\exp\left
(-\tilde {\cal A}_{22}t\right).\label{5.19}\end{equation}
Then Eq. (\ref{5.17}) assumes the form ($x=t-\tau$)
\begin{equation}{\bf D}=\mbox{sym}\left\{\int_0^t\exp\left({\cal A}_{
11}x\right){\cal A}_{12}\exp\left(-{\cal A}_{22}x\right)\,\mbox{d}
x\,{\bf F}\tilde {\cal A}_{12}\right\}.\label{5.20}\end{equation}

The explicit expressions for the matrix elements of $2\times 2$ matrix $\mu$
of (\ref{5.12}) are as follows:
\begin{equation}\mu_{11}=\frac 12\sum_i\left[-\Delta_iS_i^{(+)}-\omega_
0^{-1}G_iS_i^{(-)}+\kappa_iC_i^{(+)}\right],\label{5.23}\end{equation}
\begin{equation}\mu_{12}=\frac 12\sum_i\left[\omega_0\kappa_iS_i^{
(-)}+\omega_0\Delta_iC_i^{(-)}+G_iC_i^{(+)}\right],\label{5.24}\end{equation}
\begin{equation}\mu_{21}=\frac 12\sum_i\left[\omega_0^{-1}\kappa_
iS_i^{(-)}-\omega_0^{-1}\Delta_iC_i^{(-)}-Z_iC_i^{(+)}\right],\label{5.25}\end{equation}
\begin{equation}\mu_{22}=\frac 12\sum_i\left[-\Delta_iS_i^{(+)}-\omega_
0Z_iS_i^{(-)}-\kappa_iC_i^{(+)}\right].\label{5.26}\end{equation}
We have introduced the notation
\begin{equation}S_i^{(\pm )}=\frac {\sin(\omega_i-\omega_0)t}{\omega_
i-\omega_0}\pm\frac {\sin(\omega_i+\omega_0)t}{\omega_i+\omega_0}
,\label{5.27}\end{equation}
\begin{equation}C_i^{(\pm )}=\frac {1-\cos(\omega_i-\omega_0)t}{\omega_
i-\omega_0}\pm\frac {1-\cos(\omega_i+\omega_0)t}{\omega_i+\omega_
0}.\label{5.28}\end{equation}
Other parameters are the bilinear combinations of the coupling 
constants:
\begin{equation}\begin{array}{ccc}
\Delta_i=g_iz_i-u_iv_i,&\qquad&\kappa_i=\omega_iz_iv_i+g_iu_i/\omega_
i,\\
G_i=\omega_iv_i^2+g_i^2/\omega_i,&\qquad&Z_i=\omega_iz_i^2+u_i^2/
\omega_i.\end{array}
\label{5.29}\end{equation}

Eq.  (\ref{5.20}) results in the following expressions for the
elements of the diffusion matrix ${\bf D}$:
\begin{equation}D_{11}=\frac 12\sum_i\omega_if_i\left[G_iS_i^{(+)}
+\omega_0\Delta_iS_i^{(-)}-\omega_0\kappa_iC_i^{(-)}\right],\label{5.31}\end{equation}
\begin{equation}D_{22}=\frac 12\sum_i\omega_if_i\left[Z_iS_i^{(+)}
+\omega_0^{-1}\Delta_iS_i^{(-)}+\omega_0^{-1}\kappa_iC_i^{(-)}\right
],\label{5.32}\end{equation}
\begin{equation}D_{12}=\frac 12\sum_i\omega_if_i\left[-\kappa_iS_
i^{(+)}+(2\omega_0)^{-1}\left(\omega_0^2Z_i-G_i\right)C_i^{(-)}\right
].\label{5.33}\end{equation}

We see that in the general case both the drift matrix and the diffusion
matrix have rather complicated time dependences.  However, under
certain conditions the formulas can be simplified, if we proceed to 
the {\em continuum limit}.  This means that we assume the number of
oscillators in the reservoir to be very large and the frequencies
$\omega_i$ to be so close to each other that we may replace the sums by
integrals over $\mbox{d}\omega_i\equiv\mbox{d}\omega$.  Then we need
to calculate integrals of the following type:
\begin{equation}\sigma_{\pm}=\int\varphi (\omega )\frac {\sin(\omega
\pm\omega_0)t}{\omega\pm\omega_0}\,\mbox{d}\omega ,\label{5.34}\end{equation}
\begin{equation}\sigma_{\pm}=\int\varphi (\omega )\frac {1-\cos(\omega
\pm\omega_0)t}{\omega\pm\omega_0}\,\mbox{d}\omega .\label{5.35}\end{equation}
From physical considerations it is clear that the elements of the
matrices $\mu$ and ${\bf D}$ are determined mainly by the terms in Eqs.
(\ref{5.23})-(\ref{5.26}) and (\ref{5.31})-(\ref{5.33}) that correspond
to frequencies near $\omega_0$, since the most effective interactions
between the oscillator under study and the thermostat oscillators 
take place under resonance condition.  Indeed, for a sufficiently
smooth function $\varphi (\omega )$ and for $t\gg\omega_0^{-1}$,
only points belonging to
the domain $|\omega -\omega_0|\le t^{-1}$ make
a significant contribution to the
integrals (\ref{5.34}) and (\ref{5.35}), due to the rapid oscillations of the
trigonometric functions outside this domain.  Thus, assuming that
$\omega =\omega_0$ in all the functions except $\sin(\omega +\omega_0)t$,
we may evaluate the
integral $\sigma_{+}$ as follows:
\[\sigma_{+}\approx\frac {\varphi (\omega_0)}{2\omega_0}\int_{\omega_
1}^{\omega_2}\sin(\omega +\omega_0)t\,\mbox{d}\omega =\frac {\varphi 
(\omega_0)}{2\omega_0t}\left[\cos(\omega_1+\omega_0)t-\cos(\omega_
2+\omega_0)t\right].\]
Consequently, we may neglect the value of $\sigma_{+}$ at $\omega_
0t\gg 1$. The same is true for the integral
\[\int\varphi (\omega )\frac {\cos(\omega +\omega_0)t}{\omega +\omega_
0}\,\mbox{d}\omega\sim {\cal O}\left(\frac 1{\omega_0t}\right).\]
As concerns the integral $\sigma_{-}$, its value does not depend on time for
$t\gg\omega_0^{-1}$, since making the substitutions
$x=\omega -\omega_0$, $y=xt$, we obtain
\begin{equation}\sigma_{-}=\varphi (\omega_0)\int\frac {\sin xt}x\,\mbox{d}
x=\varphi (\omega_0)\int\frac {\sin y}y\,\mbox{d}y=\pi\varphi (\omega_
0).\label{5.36}\end{equation}
To evaluate the integral $\int\varphi (\omega )\frac {\cos(\omega 
-\omega_0)t}{\omega -\omega_0}\,\mbox{d}\omega$ we use the Taylor 
expansion $\varphi (\omega )=\varphi (\omega_0)+\varphi'(\omega_0
)(\omega -\omega_0)+\cdots$, make the substitution  
$x=\omega -\omega_0$, and expand the limits of integration from $
-\infty$ to $\infty$. 
Then the first integral vanishes, because the function $\cos(xt)/x$
is odd. The second integral decreases at least as $1/t$ when $t\to
\infty$. 
Therefore, we may assume that the integrals $\delta_{\pm}$ do not
depend on time at $\omega_0t\gg 1$:
\begin{equation}\delta_{\pm}=\int\frac {\varphi (\omega )}{\omega 
-\omega_0}\,\mbox{d}\omega .\label{5.37}\end{equation}
In the case of the ``minus'' sign the principal value of the integral 
is implied: it is designated with the symbol $\int^{\prime}$. We arrive
at the following expressions for the matrix elements of the matrices
$\mu$ and ${\bf D}$ in the continuum limit:
\begin{equation}\mu_{11}=-\frac 12\pi\nu (\omega_0)\left[\Delta (
\omega_0)+\omega_0^{-1}G(\omega_0)\right]+\int^{\prime}\frac {\omega\kappa 
(\omega )\nu (\omega )}{\omega^2-\omega_0^2}\,\mbox{d}\omega ,
\label{5.38}\end{equation}
\begin{equation}\mu_{12}=\frac 12\pi\nu (\omega_0)\omega_0\kappa 
(\omega_0)+\int^{\prime}\frac {\nu (\omega )}{\omega^2-\omega_0^2}\left
[\omega_0^2\Delta (\omega )+\omega G(\omega )\right]\,\mbox{d}\omega 
,\label{5.39}\end{equation}
\begin{equation}\mu_{21}=\frac 12\pi\nu (\omega_0)\omega_0^{-1}\kappa 
(\omega_0)-\int^{\prime}\frac {\nu (\omega )}{\omega^2-\omega_0^2}\left
[\Delta (\omega )+\omega Z(\omega )\right]\,\mbox{d}\omega ,\label{5.40}\end{equation}
\begin{equation}\mu_{22}=-\frac 12\pi\nu (\omega_0)\left[\Delta (
\omega_0)+\omega_0Z(\omega_0)\right]-\int^{\prime}\frac {\omega\kappa 
(\omega )\nu (\omega )}{\omega^2-\omega_0^2}\,\mbox{d}\omega ,\label{5.41}\end{equation}
\begin{equation}D_{11}=\frac 12\pi\nu (\omega_0)f(\omega_0)\omega_
0\left[\omega_0\Delta (\omega_0)+G(\omega_0)\right]-\omega_0^2
\int^{\prime}\frac {\omega f(\omega )\kappa (\omega )\nu (\omega )}{\omega^
2-\omega_0^2}\,\mbox{d}\omega ,\label{5.42}\end{equation}
\begin{equation}D_{22}=\frac 12\pi\nu (\omega_0)f(\omega_0)\left[
\Delta (\omega_0)+\omega_0Z(\omega_0)\right]+\int^{\prime}\frac {\omega 
f(\omega )\kappa (\omega )\nu (\omega )}{\omega^2-\omega_0^2}\,\mbox{d}
\omega ,\label{5.43}\end{equation}
\begin{equation}D_{12}=-\frac 12\pi\nu (\omega_0)\omega_0f(\omega_
0)\kappa (\omega_0)+\int^{\prime}\frac {\omega f(\omega )\nu (\omega 
)}{\omega^2-\omega_0^2}\left[\omega_0^2Z(\omega )-G(\omega )\right
]\,\mbox{d}\omega .\label{5.44}\end{equation}
Here $\nu (\omega )$ is the {\em density of states\/} function,
whereas the functions
$Z(\omega )$, $G(\omega )$, etc.  are obvious generalizations of functions
defined in Eq.  (\ref{5.29}).

In principle, the frequency dependences of the coupling constants can be
chosen in such a way that the integrals in  Eqs.
(\ref{5.38})-(\ref{5.44}) vanish. For instance, this is possible
provided the corresponding combinations of functions  
$\kappa$, $\nu$, $f$, $G$, $Z$, $\Delta /\omega$, understood as functions
of the argument
$x=\omega^2$, do not change their values under the reflection in the point
$x_0=\omega_0^2$, and these functions (or at least the density of states) 
decrease sufficiently rapidly with distance from the point $x_0$ in
both directions. In such a case all the coefficients $\mu_{ik}$ and
$D_{ik}$ are determined by the values of the aforementioned functions
at the point $\omega_0$.  Moreover, the diffusion coefficients are
proportional to the corresponding elements of the matrix $\mu$:
\begin{equation}D_{11}=-\omega_0^2f_0\mu_{11},\qquad D_{22}=-f_0\mu_{
22},\qquad D_{12}=-f_0\mu_{12}=-\omega_0^2f_0\mu_{21}.\label{5.45}\end{equation}

We see that in the continuous weak coupling limit the reduced 
Wigner function of the oscillator obeys (under certain conditions) 
the Fokker-Planck equation (\ref{2.22}) {\em with time-independent 
coefficients\/} at times $t\gg\omega_0^{-1}${\em .\/}  Let us check,
however, whether
condition (\ref{2.38}) is fulfilled.  Since $\mbox{Tr}{\bf A}=\mbox{Tr}
\mu$, then due to Eq.
(\ref{5.45}) we must check the inequality
\begin{equation}\omega_0^2f_0^2\mu_{11}\mu_{22}-f_0^2\mu_{12}^2\ge
\hbar^2\left(\mu_{11}+\mu_{22}\right)^2/16.\label{5.46}\end{equation}
Taking into account Eqs. (\ref{5.38})-(\ref{5.41}) and Eq. (\ref{5.29}),
we arrive at the inequality (all the functions are taken at the point
$\omega =\omega_0$)
\begin{equation}(4f\omega /\hbar )^2\left[2\Delta^2+\Delta (\omega 
Z+G/\omega )\right]\ge\left[2\Delta +(\omega Z+G/\omega )\right]^
2.\label{5.47}\end{equation}

It cannot be satisfied for an arbitrary choice of coupling constants. 
For instance, it is violated if $\Delta =0$, or when any three of four
coefficients $g$, $z$, $u$, $v$ vanish.

This result seems paradoxical.  Indeed, we started from the exact 
equation of motion for the density matrix of a closed system, 
found the exact solution to this equation, and after this we 
performed averaging over the thermostat degrees of freedom.  
Since the laws of quantum mechanics were not violated at any
step, the reduced density matrix {\em must\/} be positive definite at any
time {\bf for quite arbitrary coupling constants}.  On the other hand, if, 
for instance, $z_i=u_i=v_i=0$, then inequality (\ref{5.47}) does not 
hold.  Hence, following the reasonings given in Sec. \ref{sec2},
we could obtain a nonpositive definite density matrix
in the process of evolution!

This apparent contradiction is resolved in the following way.  
Inequality (\ref{2.38}) is a necessary and sufficient condition
ensuring that {\bf any} density matrix that was positive definite at
{\bf any} instant of time will remain positive definite at all
subsequent moments.  But in the problem under study we have the 
{\bf selected} instant $t=0$:  this is just the moment when the 
interaction with the thermostat was turned on.  Since in the 
presence of the interaction with the environment the evolution of 
the oscillator density matrix is {\em nonunitary}, the set of density 
matrices $\rho (t)$ arising from all initially admissible density matrices 
$\rho (0)$ {\bf does not coincide} with the set of all admissible density 
matrices.  In particular, correct initial density matrices {\bf cannot }
turn into the specific ones that could become nonpositive
definite at some instant of time in the case of violation of 
inequality (\ref{5.47}). Therefore, there is no need to check
conditions like (\ref{2.38}) or (\ref{5.46}), (\ref{5.47}) when the 
density matrix of the subsystem is obtained by
reduction of the {\bf exact} density matrix of the closed system: the 
reduced density matrix (calculated with the proper accuracy) turns 
out to be positive definite {\bf automatically}.

However, if the goal is the derivation of a self-consistent 
Fokker-Planck equation on the basis of a ``microscopic'' model of 
the oscillator interacting with a large reservoir, then we have to 
recognize that the underlying ``microscopic'' model cannot be quite 
arbitrary:  its parameters must satisfy rather strong restriction 
(\ref{5.47}), in order to prevent the appearance of the unphysical 
solutions when this equation is applied to arbitrary initial states.  

Returning to the analysis of Eqs. (\ref{5.38})-(\ref{5.45}), we notice
that the matrix
\begin{equation}{\bf F}_0=\left\Vert\begin{array}{cc}
\omega_0^2f_0&0\\
0&f_0\end{array}
\right\Vert\label{5.48}\end{equation}
satisfies, due to Eq. (\ref{5.45}), the relation
\begin{equation}{\bf A}{\bf F}_0+{\bf F}_0\widetilde{{\bf A}}+2{\bf D}
=0.\label{5.49}\end{equation}
This means that ${\bf F}_0$ is the {\em steady state solution\/} to Eq. (\ref{2.23})
for the oscillator variance matrix, {\bf independently} of the concrete
values of the coefficients of the drift matrix $\mu_{ik}$. In particular, if
\begin{equation}f_0=\frac {\hbar}{2\omega_0}\coth\left(\frac {\hbar
\omega_0}{2kT}\right),\label{5.50}\end{equation}
then the matrix (\ref{5.48}) coincides with the equilibrium
variance matrix of the oscillator.  Consequently, the steady state
solution of the
Fokker-Planck equation with the coefficients (\ref{5.38})-(\ref{5.44}) is
the {\em equilibrium Wigner function of the oscillator}, if the thermostat
itself is in the equilibrium state.  Moreover, we may imagine a
situation with 
\begin{equation}f_0=\frac {\hbar}{2\omega_0}\sum_{i=0}^n\alpha_i\coth\left
(\frac {\hbar\omega_0}{2kT_i}\right),\qquad\sum_{i=1}^n\alpha_i=1
,\qquad\alpha_i\ge 0,\label{5.51}\end{equation}
Then, the oscillator under study exhibits relaxation to the Gaussian
steady state with the variance matrix given by Eqs.  (\ref{5.48})
and (\ref{5.51}).  Such a situation may be realized when the 
thermostat consists of several large independent subsystems
possessing their own temperatures (see, e.g.,  \cite{10}).  In this 
case $n$ is the number of subsystems, and $\alpha_i$ is the ``weight'' of
each subsystem.  

We see that the final steady state of the oscillator weakly 
interacting with a thermostat does not depend on the concrete
values of the coupling constants, provided the thermostat was initially
in an equilibrium or quasi-equilibrium (described by Eq.
(\ref{5.51}) state.  Now let us analyze possible forms of the
drift matrix ${\bf A}$.
Due to the property $\mu_{12}=\omega_0^2\mu_{21}$
its characteristic equation $\det({\bf A} -\lambda {\bf I})=0$ reads
\begin{equation}\lambda^2-\left(\mu_{11}+\mu_{22}\right)\lambda +
\omega_0^2+\mu_{11}\mu_{22}-\mu_{12}\mu_{21}=0.\label{5.52}\end{equation}
The solutions to this equations can be written as $\lambda_{1,2}=
-\gamma\pm i\omega_{*}$, 
with
\begin{equation}\omega_{*}=\left[\omega_0^2-\frac 14\left(\mu_{11}
-\mu_{22}\right)^2-\mu_{12}\mu_{21}\right],\label{5.53}\end{equation}
\begin{equation}\gamma =-\frac 12\mbox{Tr}\mu =-\frac 12\left(\mu_{
11}+\mu_{22}\right)=\frac 14\pi\nu (\omega_0)\left[2\Delta_0+\omega_
0Z_0+\omega_0^{-1}G_0\right].\label{5.54}\end{equation}
The last formula holds even when the integrals in Eqs. (\ref{5.38})
and (\ref{5.41}) are not equal to zero. Assuming the damping coefficient
$\gamma$ to be positive and comparing Eqs. (\ref{5.54}) and (\ref{5.47}), we
may rewrite inequality (\ref{5.47}) as follows (recall that all
functions are taken at $\omega =\omega_0$):
\begin{equation}(4f\omega /\hbar )^2\Delta\ge 2\Delta +(\omega Z+
G/\omega )=4\gamma /[\pi\nu (\omega )].\label{5.55}\end{equation}
In particular, for an equilibrium thermostat at zero temperature
(when $f=\hbar /2\omega$) we obtain the relation
\begin{equation}2\Delta\ge\omega Z+G/\omega ,\label{5.56}\end{equation}
which due to Eq. (\ref{5.29}) turns into the inequality
\begin{equation}(u+v)^2+(\omega z-g/\omega )^2\le 0.\label{5.57}\end{equation}

Thus we arrive at a striking conclusion:  the microscopic model
of the interaction between the oscillator and the equilibrium 
reservoir at zero temperature, based on the general quadratic 
Hamiltonian (\ref{5.1}), leads to the Fokker-Planck equation with 
time independent coefficients, valid for any physically admissible 
initial states of the oscillator, if and only if the coupling constants 
at the main oscillator frequency satisfy the restriction
\begin{equation}u_0=-v_0,\qquad z_0=g_0/\omega_0^2.\label{5.58}\end{equation}
In this case Eqs. (\ref{5.38})-(\ref{5.41}) yield
\begin{equation}\mu_{11}=\mu_{22},\qquad\mu_{12}=\mu_{21}=0,\label{5.60}\end{equation}
so we have the unique drift matrix
\begin{equation}\mu =\left\Vert\begin{array}{cc}
-\gamma&0\\
0&-\gamma\end{array}
\right\Vert\label{5.59}\end{equation}
with the damping coefficient
\begin{equation}\gamma =\pi\nu (\omega_0)\Delta_0,\qquad\Delta_0=
u_0^2+g_0^2/\omega_0^2.\label{5.61}\end{equation}
Due to Eq. (\ref{5.60}) the frequency $\omega_{*}$ in formula (\ref{5.53})
{\em exactly\/} equals the oscillator eigenfrequency $\omega$$_0$.

The equations of motion for the average values of the coordinate 
and momentum read
\begin{equation}\dot {p}=-\gamma p-\omega_0^2x,\label{5.62}\end{equation}
\begin{equation}\dot {x}=p-\gamma x.\label{5.63}\end{equation}
One can see that consistent quantum mechanical consideration do
not result in the conventional classical equations (\ref{2.20}).  To 
understand the origin of Eqs. (\ref{5.62}) and (\ref{5.63}), let us
introduce the annihilation and creation operators
\begin{equation}\hat {a}=\frac {\omega_0\hat {x}+i\hat {p}}{\sqrt {
2\hbar\omega_0}},\qquad\hat {a}^{\dagger}=\frac {\omega_0\hat {x}
-i\hat {p}}{\sqrt {2\hbar\omega_0}},\label{5.64}\end{equation}
which are the most natural for the description of a quantum 
oscillator. It turns out that precisely
Eqs. (\ref{5.62}), (\ref{5.63}) lead to
{\em uncoupled\/} equations for $\hat {a}$ and $\hat {a}^{\dag}$:
\begin{equation}\dot{\hat {a}}=-i\omega_0\hat {a}-\gamma\hat {a},
\qquad\dot{\hat {a}}^{\dag}=i\omega_0\hat {a}^{\dag}-\gamma\hat {
a}^{\dag}.\label{5.65}\end{equation}

Denoting the annihilation and creation operators for the
thermostat oscillators by $\hat {c}_i$ and $\hat {c}_i^{\dag}$ we may
rewrite the
interaction Hamiltonian (\ref{5.1}) (for thermostat oscillators
whose frequencies coincide with $\omega_0$) as follows:
\begin{equation}\hat {H}_{\mbox{int}}(\omega_0)=\hbar\sum_{\omega_
i=\omega_0}\left(\delta_0\hat a\hat c_i^{\dag}+\delta_0^{*}\hat a^{
\dag}\hat c_i\right),\label{5.66}\end{equation}
\[\delta_0=g_0/\omega_0+iu_0,\qquad\Delta_0=|\delta_0|^2.\]

Hamiltonian (\ref{5.66}) was considered in almost every paper
devoted to the models of a quantum damped oscillator (see, e.g.,
review \cite{Dekker}).  But frequently it was chosen only because it is
the simplest one.  We have shown in fact that it is the only
possible quadratic interaction Hamiltonian ensuring the
validity of the resulting Fokker-Planck equation for any initial 
states taken at any initial times.  This is probably related to the
quantum nature of the interactions between the systems:  each
act of interaction must consist in the annihilation of a quantum in 
one system and its creation in another system.  Precisely the Hamiltonian
(\ref{5.66}) expresses this property in the most distinct form.

Nonetheless we have no sufficient grounds for claiming that other
interaction (quadratic) Hamiltonians should be excluded. But they 
will result in the Fokker-Planck equation with time-dependent
drift and diffusion matrices describing nonexponential relaxation.

Recall that we assumed the integrals in Eqs.
(\ref{5.38})-(\ref{5.44}) to equal zero. What will happen if we 
abandon this assumption, but assume instead that the constraints  
(\ref{5.58}) hold for {\bf all} the coupling constants?
In this case we have
$\kappa (\omega )\equiv 0$ at all frequencies. Furthermore, $G(\omega
)\equiv\omega^2Z(\omega )\equiv\omega\Delta (\omega )$. 
Consequently, the integral terms ``survive'' only in the off-diagonal 
elements of both matrices $\mu$ and ${\bf D}$:
\[\mu_{11}=\mu_{22}=-\gamma ,\]
\[\mu_{12}=\int^{\prime}\frac {\nu (\omega )\Delta (\omega )}{\omega^
2-\omega_0^2}\left[\omega_0^2+\omega^2\right]\,\mbox{d}\omega ,\]
\[\mu_{21}=-\int^{\prime}\frac {\nu (\omega )\Delta (\omega )}{\omega^
2-\omega_0^2}\,\mbox{d}\omega ,\]
\begin{equation}D_{11}=\omega_0^2D_{22},\qquad D_{22}=\gamma f(\omega_
0),\qquad D_{12}=-\frac 1{\pi}\int f(\omega )\gamma (\omega )\,\mbox{d}
\omega\label{5a}\end{equation}
(the function $\gamma (\omega )$ is defined by Eq. (\ref{5.61}) with $
\omega_0$ replaced by $\omega$).

The presence of nonzero coefficients $\mu_{12}$ and $\mu_{21}$ may be 
interpreted as some kind of renormalization of mass and 
eigenfrequency of the main oscillator due to the interaction with the
environment.  However, inequality (\ref{2.37}) is obviously violated 
at zero temperature, when $f=\hbar /2\omega$, since due to Eq.  (\ref{5a})
coefficient $D_{12}$ is strictly negative for all temperatures (evidently, 
both functions $\nu (\omega )$ and $f(\omega )$ are positive).
This example shows
once more that the self-consistent Fokker-Planck equations with 
time-independent coefficients can be derived from microscopic
models only in exceptional cases.

\setcounter{equation}{0}
\section{Oscillator in a magnetic field.  Weak coupling with a 
thermostat} 

We now consider, within the framework of the same scheme, the
case where the subsystem under study is a two-dimensional
isotropic oscillator with eigenfrequency $\omega_0$ and mass $m$ placed 
in a uniform magnetic field ${\cal H}$ characterized by the cyclotron
frequency 
\begin{equation}\omega =e{\cal H}/mc.\label{6.1}\end{equation}
The Hamiltonian of this subsystem reads
\begin{equation}\hat {H}_0=\frac 1{2m}\left(\hat\pi_x^2+\hat\pi_y^
2\right)+\frac 12m\omega_0^2\left(x^2+y^2\right),\label{6.2}\end{equation}
where $\hat{\pi}_x$ and $\hat{\pi}_y$ are the operators of the
{\em kinetic\/} momentum
projections, related to the {\em canonical\/} momentum ${\bf p}$ and the
vector potential ${\bf A}$ in the usual way:
\begin{equation}\pi ={\bf p}-(e/c){\bf A}(x,y),\qquad\left[\hat\pi_
x,\hat\pi_y\right]=i\hbar m\omega .\label{6.3}\end{equation}

The problem of constructing the Fokker-Planck equation for this 
subsystem is reduced to that solved in the previous section, 
because the Hamiltonian (\ref{6.2}) can be expressed as a sum of two
oscillator Hamiltonians:
\begin{equation}\hat {H}_0=\hbar\omega_{+}\left(\hat a^{\dag}\hat 
a+\frac 12\right)+\hbar\omega_{-}\left(\hat b^{\dag}\hat b+\frac 
12\right).\label{6.4}\end{equation}
The annihilation operators
\begin{equation}\hat {a}=(2m\hbar\Omega )^{-1/2}\left[\hat\pi_x+i
\hat\pi_y+m\omega_{-}\left(\hat y-i\hat x\right)\right],\label{6.5}\end{equation}
\begin{equation}\hat {b}=(2m\hbar\Omega )^{-1/2}\left[\hat\pi_x-i
\hat\pi_y-m\omega_{+}\left(\hat y+i\hat x\right)\right]\label{6.6}\end{equation}
satisfy the commutation relations
\begin{equation}\left[\hat a,\hat a^{\dag}\right]=\left[\hat b,\hat 
b^{\dag}\right]=1,\qquad\left[\hat a,\hat b\right]=\left[\hat a,\hat 
b^{\dag}\right]=0.\label{6.7}\end{equation}
The frequencies are defined as follows:
\begin{equation}\omega_{\pm}=\frac 12(\Omega\pm\omega ),\qquad\Omega 
=\left(\omega^2+4\omega_0^2\right)^{1/2},\qquad\omega_{+}\omega_{
-}=\omega_0^2.\label{6.8}\end{equation}

In the continuous weak coupling limit the self-consistent equations 
of motion for the first-order average values of the operators $\hat {a}$
and $\hat {b}$ are given by Eq. (\ref{5.65}), provided one replaces
$\omega_0$ with $\omega_{+}$ and $\omega_{-}$.
Furthermore, two different damping coefficients are
possible: they are determined by the density of states and coupling 
constants at the frequencies $\omega_{\pm}$ (see Eqs. (\ref{5.61}) and
(\ref{5.66})):
\begin{equation}\gamma_{\pm}=\pi\nu (\omega_{\pm})|\delta (\omega_{
\pm})|^2.\label{6.8prime}\end{equation}
The relations inverse to Eqs. (\ref{6.5}) and (\ref{6.6}) read
\begin{equation}\hat{\pi}_x=(m\hbar /2\Omega )^{1/2}\left[\omega_{
+}\left(\hat a+\hat a^{\dag}\right)+\omega_{-}\left(\hat b+\hat b^{
\dag}\right)\right],\label{6.9}\end{equation}
\begin{equation}\hat{\pi}_y=i(m\hbar /2\Omega )^{1/2}\left[\omega_{
+}\left(\hat a^{\dag}-\hat a\right)+\omega_{-}\left(\hat b-\hat b^{
\dag}\right)\right],\label{6.10}\end{equation}
\begin{equation}\hat {x}=i(\hbar /2m\Omega )^{1/2}\left[\hat a-\hat 
a^{\dag}+\hat b-\hat b^{\dag}\right],\label{6.11}\end{equation}
\begin{equation}\hat {y}=(\hbar /2m\Omega )^{1/2}\left[\hat a+\hat 
a^{\dag}-\hat b-\hat b^{\dag}\right].\label{6.12}\end{equation}
The average values of the coordinates and the kinetic momenta 
obey equations resulting from equations of the form of (\ref{5.65}):
\begin{equation}\dot{\pi}_x=-\alpha\pi_x+\omega\pi_y-m\omega_0^2x
+m\omega_0^2\epsilon y,\label{6.13}\end{equation}
\begin{equation}\dot{\pi}_y=-\omega\pi_x-\alpha\pi_y-m\omega_0^2\epsilon 
x-m\omega_0^2y,\label{6.14}\end{equation}
\begin{equation}\dot {x}=m^{-1}\pi_x-m^{-1}\epsilon\pi_y-\eta x,\label{6.15}\end{equation}
\begin{equation}\dot {y}=m^{-1}\epsilon\pi_x+m^{-1}\pi_y-\eta y.\label{6.16}\end{equation}
We have introduced the notation
\begin{equation}\alpha =\left(\gamma_{+}\omega_{+}+\gamma_{-}\omega_{
-}\right)/\Omega ,\quad\eta =\left(\gamma_{+}\omega_{-}+\gamma_{-}
\omega_{+}\right)/\Omega ,\quad\epsilon =\left(\gamma_{-}-\gamma_{
+}\right)/\Omega .\label{6.17}\end{equation}
The second-order equations of motion read
\begin{equation}\ddot {x}+\left(\gamma_{-}+\gamma_{+}\right)\dot {
x}-\omega\dot {y}+\left(\omega_0^2+\gamma_{-}\gamma_{+}\right)x-\left
(\gamma_{-}\omega_{+}-\gamma_{+}\omega_{-}\right)y=0,\label{6.18}\end{equation}
\begin{equation}\ddot {y}+\left(\gamma_{-}+\gamma_{+}\right)\dot {
y}+\omega\dot {x}+\left(\omega_0^2+\gamma_{-}\gamma_{+}\right)y+\left
(\gamma_{-}\omega_{+}-\gamma_{+}\omega_{-}\right)x=0.\label{6.19}\end{equation}

We see that ``one-photon'' interaction with a thermostat of the form of
(\ref{5.66}) results in {\bf coordinate-dependent forces\/}
perpendicular to the vector ${\bf r}=(x,y)$ and proportional to the damping
coefficients.  The necessity of introducing such forces was shown 
earlier in \cite{30,37,38} within the framework of a phenomenological
approach.  As was demonstrated in these papers, if the second-order
equations of motion (for a charged particle or oscillator placed in a 
uniform magnetic field) contain the term $-\gamma\dot {{\bf r}}$,
then it is impossible
to satisfy simultaneously Eq.  (\ref{2.23}) with the equilibrium
matrix ${\cal M}(T)$ for all temperatures (including $T=0$) and condition 
(\ref{2.24}), unless a force of the form ${\bf f}=[{\bf h}\times {\bf r}]$
is introduced.  Now we
have arrived at the same result on the base of the microscopic 
model.  Moreover, the relation between the velocity and the 
kinetic momentum becomes much more complicated than in
the conservative case (see Eqs.  (\ref{6.15}) and (\ref{6.16}).

It is clear from the preceding section that the steady state Wigner 
function at $t\to\infty$ coincides with the equilibrium distribution that
was found in refs. \cite{8,15}. Since this distribution is Gaussian, it 
is completely determined (see Eq. (\ref{3.13})) by the equilibrium
variance matrix
\begin{equation}{\cal M}^{(\mbox{eq})}=\left\Vert\begin{array}{cccc}
{\cal M}_{\pi}&0&0&{\cal M}_a\\
0&{\cal M}_{\pi}&-{\cal M}_a&0\\
0&-{\cal M}_a&{\cal M}_{\rho}&0\\
{\cal M}_a&0&0&{\cal M}_{\rho}\end{array}
\right\Vert ,\label{6.20}\end{equation}
\begin{equation}{\cal M}_{\pi}=\frac {m\hbar\Omega}{4Q}\left[\left
(1+\frac {\omega^2}{\Omega^2}\right)\sinh\tilde\Omega -2\frac {\omega}{
\Omega}\sinh\tilde\omega\right],\label{6.21}\end{equation}
\begin{equation}{\cal M}_{\rho}=\frac {\hbar\sinh\tilde{\Omega}}{
m\Omega Q},\label{6.22}\end{equation}
\begin{equation}{\cal M}_a=\frac {\hbar}{2Q}\left[\frac {\omega}{
\Omega}\sinh\tilde\Omega -\sinh\tilde\omega\right],\label{6.23}\end{equation}
where
\begin{equation}Q(\beta )=\cosh\tilde{\Omega }-\cosh\tilde{\omega }
,\qquad\tilde{\Omega }=\frac 12\beta\hbar\Omega ,\qquad\tilde{\omega }
=\frac 12\beta\hbar\omega ,\label{6.24}\end{equation}
and $\beta =1/kT$ is the inverse temperature of the thermostat.

The drift matrix ${\bf A}$ corresponding to Eqs. (\ref{6.13})-(\ref{6.16})
reads
\begin{equation}{\bf A}=\left\Vert\begin{array}{cccc}
-\alpha&\omega&-m\omega_0^2&m\omega_0^2\epsilon\\
-\omega&-\alpha&-m\omega_0^2\epsilon&-m\omega_0^2\\
m^{-1}&-m^{-1}\epsilon&-\eta&0\\
m^{-1}\epsilon&m^{-1}&0&-\eta\end{array}
\right\Vert .\label{6.25}\end{equation}
Putting this matrix into Eq. (\ref{5.49}) with the matrix ${\cal M}^{
(\mbox{eq})}$ instead 
of ${\bf F}_0$ we obtain the diffusion matrix
\begin{equation}{\bf D}=\left\Vert\begin{array}{cccc}
D_{\pi}&0&0&D_a\\
0&D_{\pi}&-D_a&0\\
0&-D_a&D_{\rho}&0\\
D_a&0&0&D_{\rho}\end{array}
\right\Vert\label{6.26}\end{equation}
with the following coefficients,
\begin{equation}D_{\pi}=\frac {m\hbar}{2\Omega Q}\left[\left(\gamma_{
+}\omega_{+}^2+\gamma_{-}\omega_{-}^2\right)\sinh\tilde\Omega -\left
(\gamma_{+}\omega_{+}^2-\gamma_{-}\omega_{-}^2\right)\sinh\tilde\omega\right
],\label{6.27}\end{equation}
\begin{equation}D_a=\frac {\hbar}{2\Omega Q}\left[\left(\gamma_{+}
\omega_{+}-\gamma_{-}\omega_{-}\right)\sinh\tilde\Omega -\left(\gamma_{
+}\omega_{+}+\gamma_{-}\omega_{-}\right)\sinh\tilde\omega\right],\label{6.28}\end{equation}
\begin{equation}D_{\rho}=\frac {\hbar}{2m\Omega Q}\left[\left(\gamma_{
+}+\gamma_{-}\right)\sinh\tilde\Omega -\left(\gamma_{+}-\gamma_{-}\right
)\sinh\tilde\omega\right].\label{6.29}\end{equation}
In particular, at zero temperature $(\beta =\infty )$ we obtain
\begin{eqnarray}
D_{\pi}^{(\mbox{low})}&=&\frac {m\hbar}{2\Omega}\left(\gamma_{+}\omega_{
+}^2+\gamma_{-}\omega_{-}^2\right),
\nonumber \\
D_a^{(\mbox{low})}&=&\frac {\hbar}{2\Omega}\left(\gamma_{+}\omega_{
+}-\gamma_{-}\omega_{-}\right),
\nonumber \\
D_{\rho}^{(\mbox{low})}&=&\frac {\hbar}{2m\Omega}\left(\gamma_{+}
+\gamma_{-}\right).
\end{eqnarray}
In the opposite, high-temperature, case $(\beta\to 0)$ we have
\begin{equation}
D_{\pi}^{(\mbox{high})}=mkT\alpha ,\qquad D_{\rho}^{(\mbox{high}
)}=\frac {kT\eta}{m\omega_0^2},\qquad D_a^{(\mbox{high})}=-kT\epsilon 
.\label{6.31}
\end{equation}
It is noteworthy that all three diffusion coefficients remain
nonzero even in the high temperature limit, which is usually identified
with the quasiclassical limit. Recall that in classical
statistical mechanics it is usually implied that the only nonzero
diffusion coefficient is $D_{\pi}$.

Various sets of the diffusion coefficients compatible with
inequality (\ref{2.24}) and leading to an equilibrium steady state
(with the variance matrix (\ref{6.20})) in the limit of infinitely 
small damping were constructed within the framework of the
phenomenological approach in \cite{30}.  However, none of
them had a structure similar to that given by Eqs.
(\ref{6.27})-(\ref{6.29}).  

For example, only the coefficient $D_{\pi}$ was proportional to the
temperature in the high-temperature limit, whereas the other 
diffusion coefficients decreased as $1/kT$, in contrast to
Eq. (\ref{6.31}). This difference is due to at least two causes.
First, it was assumed in  \cite{30,37,38} that the elements $A_{41}$ and
$A_{32}$ of the drift matrix must be zero, although the elements $A_{14}$
and $A_{23}$ could be nonzero. Eq. (\ref{6.25}) shows that within the
framework of the microscopic approach such a choice is 
impossible, since all these coefficients are proportional to the
parameter $\epsilon$. Furthermore, in the aforementioned papers
we admitted the
possibility that some coefficients of the drift matrix (those related 
to the damping) could depend on temperature. In principle, such a 
possibility (i.e., the time dependence of the coupling constants in 
the interaction Hamiltonian) is not excluded within the framework of
the microscopic approach as well. Then the high temperature limit
of the diffusion matrix coefficients could be quite different from Eq.
(\ref{6.31}).

The expressions (\ref{6.25})-(\ref{6.29}) are simplified in the special 
case of the equal damping coefficients, when $\gamma_{+}=\gamma_{
-}=\gamma_0$:
\begin{equation}{\bf A}^{(0)}=\left\Vert\begin{array}{cccc}
-\gamma_0&\omega&-m\omega_0^2&0\\
-\omega&-\gamma_0&0&-m\omega_0^2\\
m^{-1}&0&-\gamma_0&0\\
0&m^{-1}&0&-\gamma_0\end{array}
\right\Vert ,\label{6.45}\end{equation}
\begin{equation}D_{\pi}^{(0)}=\frac {m\hbar\gamma_0}{2\Omega Q}\left
[\left(\omega^2+2\omega_0^2\right)\sinh\tilde\Omega -\omega\Omega\sinh
\tilde\omega\right],\label{6.46}\end{equation}
\begin{equation}D_a^{(0)}=\frac {\hbar\gamma_0}{2\Omega Q}\left[\omega\sinh
\tilde\Omega -\Omega\sinh\tilde\omega\right],\label{6.47}\end{equation}
\begin{equation}D_{\rho}^{(0)}=\frac {\hbar\gamma_0}{m\Omega Q}\sinh
\tilde{\Omega }.\label{6.48}\end{equation}

If coefficient $\gamma_{-}$ tends to zero sufficiently rapidly as $
\omega_{-}\to 0$, then 
the set of diffusion coefficients (\ref{6.27})-(\ref{6.29}) possesses 
the finite limit for a free particle in a magnetic field, when 
$\omega_0=\omega_{-}=0$, $\Omega =\omega$, $\gamma_{+}(\omega_{+}
\equiv\omega )=\gamma$:  
\begin{equation}D_{\pi}=\frac 12\gamma m\hbar\omega\coth\tilde{\omega }
,\label{6.61}\end{equation}
\begin{equation}D_a=\frac 12\gamma\hbar\coth\tilde{\omega },\label{6.62}\end{equation}
\begin{equation}D_{\rho}=\frac {\gamma\hbar}{2m\omega}\coth\tilde{
\omega },\label{6.63}\end{equation}
\begin{equation}\alpha =\gamma ,\qquad\eta =0,\qquad\epsilon =-\gamma 
/\omega .\label{6.64}\end{equation}

In this case the operators $\hat {b}$ and $\hat {b}^{\dag}$ become
{\em the integrals of motion},
whose real and imaginary parts are connected with the
{\em center-of-orbit\/} operators in a uniform magnetic field \cite{43}-\cite{45}.

One should remember, however, that the results obtained in this 
section can be justified only under rather strong limitations 
imposed on the interaction Hamiltonian.  First, it must be written 
in the specific form (\ref{5.66}) at the resonant frequencies.  
Secondly, the off-resonance terms must ensure the disappearance 
of the integral terms in Eqs.  (\ref{5.38})-(\ref{5.44}).  In
particular, the density of states must decrease sufficiently rapidly 
as $\omega\to\infty$.

All these conditions are violated, for example, in the case where
the role of a reservoir is played by a quantized electromagnetic
field coupled to the oscillator by means of the standard interaction 
Hamiltonian in the dipole approximation:
\begin{equation}\hat {H}_{\mbox{int}}=-\frac em(2\pi\hbar )^{1/2}
\sum_{j,\sigma}\frac {\left(\hat\pi\tau_{j,\sigma}\right)}{\sqrt {
\omega_j}}\left[\hat c({\bf k}_j,\sigma )+\hat c^{\dag}({\bf k}_j
,\sigma )\right],\label{6.49}\end{equation}
where $\hat {c}({\bf k}_j,\sigma )$ is the operator of annihilation of a
photon with wave vector ${\bf k}_j$, frequency $\omega_j=c|{\bf k}_
j|$, and polarization 
$\sigma$; $\tau_{j,\sigma}$ is the unit polarization vector
perpendicular to the vector ${\bf k}_j$.
The density of states is proportional to $\omega^2$ in this case, and
the integrals in Eqs.  (\ref{5.38})-(\ref{5.44}) {\em diverge}.
Consequently, in this case the radiation damping leads to
{\em nonexponential relaxation.  }

\end{document}